\documentclass[10pt]{article}
\usepackage{amsmath}
\usepackage{graphicx}
\usepackage[inline]{enumitem}
\usepackage{natbib}
\usepackage{url} 
\usepackage{bbm}
\usepackage{bm}
\usepackage{amsfonts}
\usepackage{amsthm}
\usepackage{authblk}
\usepackage{hyperref}
\usepackage[verbose=true]{geometry}

\newgeometry{
    textheight=9in,
    textwidth=5.5in,
    top=1in,
    headheight=12pt,
    headsep=25pt,
    footskip=30pt
}

\newcommand{\E}{E}
\newcommand{\Ehat}{\hat{E}}
\newcommand{\Cov}{\mathrm{cov}}
\newcommand{\Var}{\mathrm{var}}
\newcommand{\indep}{\perp\!\!\!\perp}
\newcommand{\n}[2]{\mathcal{N}\left( #1,#2\right)}
\graphicspath{{plots/}}

\DeclareMathOperator*{\argmin}{arg\,min}

\newtheorem{theorem}{Theorem}
\newtheorem{corollary}{Corollary}[theorem]
\newtheorem{lemma}{Lemma}[theorem]

\newtheorem{remark}{Remark}
\newtheorem{example}{Example}

\newcommand{\mytitle}{Optimally weighted average derivative effects}
\title{\mytitle}
\newcommand{\myappendix}{Appendix } 

\author[1]{Oliver Hines\thanks{The authors gratefully acknowledge the MRC London Intercollegiate Doctoral Training Partnership}}
\author[2]{Karla Diaz-Ordaz}
\author[3]{Stijn Vansteelandt}
\affil[1]{Department of Medical Statistics, London School of Hygiene and Tropical Medicine, London, U.K.}
\affil[2]{Department of Statistical Science, University College London, U.K.}
\affil[3]{Department of Applied Mathematics, Computer Science and Statistics, Ghent University, Ghent, Belgium}

\begin{document}
\maketitle
\abstract{Weighted average derivative effects (WADEs) are nonparametric estimands with uses in economics and causal inference. Debiased WADE estimators typically require learning the conditional mean outcome as well as a Riesz representer (RR) that characterises the requisite debiasing corrections. RR estimators for WADEs often rely on kernel estimators, introducing complicated bandwidth-dependant biases. In our work we propose a new class of RRs that are isomorphic to the class of WADEs and we derive the WADE weight that is optimal, in the sense of having minimum nonparametric efficiency bound. Our optimal WADE estimators require estimating conditional expectations only (e.g. using machine learning), thus overcoming the limitations of kernel estimators. Moreover, we connect our optimal WADE to projection parameters in partially linear models. We ascribe a causal interpretation to WADE and projection parameters in terms of so-called incremental effects. We propose efficient estimators for two WADE estimands in our class, which we evaluate in a numerical experiment and use to determine the effect of Warfarin dose on blood clotting function.}
\setlength{\parskip}{5.5pt}
\setlength{\parindent}{0pt}
\section{Introduction}
\label{introduction}

Average derivative effects (ADEs), also called average partial effects, were originally motivated for the estimation of parameters in index models \citep{Hardle1989,Powell1989,Newey1993, Imbens2009}, a problem of substantial practical interest in econometrics, with additional uses in assessing the law of total demand in economics \citep{Hardle1991}, in policy learning \citep{Athey2021}, and in causal inference \citep{Rothenhausler2019}. We consider the weighted ADE (WADE) vector, $\bm{\theta} = \E\{w(\bm{X})d\mu(\bm{X})/d\bm{X}\}$, where for an outcome $Y$ and covariate vector $\bm{X}$, we denote the conditional response function $\mu(\bm{x})=\E(Y|\bm{X}=\bm{x})$, with derivative $d\mu(\bm{x})/d\bm{x}$ and $w(\bm{x})$ is a weight function.

Estimation of WADEs is challenging even when a parametric model for $\mu$ is assumed, since, for all but the simplest parametric models, the WADE is a nonlinear function of the parameters, and some additional debiasing is required \citep{Hirshberg2018, Wooldridge2020}. Moreover, it may be difficult to correctly specify the functional form of $\mu(\bm{x})$ in practice, motivating the use of machine learning (lasso, gradient boosting, neural nets etc.) to obtain estimates $\hat{\mu}$. Estimators based on machine learning also require debiasing, since the outcome learner usually optimises a bias-variance trade off with respect to a generic loss function (mean squared error, logistic loss etc.) that does not adequately control for plug-in biases in the downstream WADE estimator \citep{Newey1993, Chernozhukov2018}. E.g. decision tree based learners are piecewise constant and therefore poorly suited to WADE estimation, even when they perform well in outcome prediction tasks.

Regardless of the outcome modelling approach, the plug-in bias is characterised by the so-called Riesz representer (RR) of $\bm{\theta}$, which is the function $\bm{l}(\bm{X})$ in the Riesz representation $\bm{\theta} = \E\{\bm{l}(\bm{X})\mu(\bm{X})\} = \E\{\bm{l}(\bm{X})Y\}$. Riesz representations are an alternative way of representing bounded, continuous and linear maps (in this case $\mu\mapsto \bm{\theta}$) the existence of which is guaranteed by Riesz's representation theorem. Under regularity assumptions, which we will describe later, \cite{Powell1989} used integration by parts to derive the RR
\begin{align}
\bm{l}(\bm{x}) = -\frac{dw(\bm{x})}{d\bm{x}} - \frac{w(\bm{x})}{f(\bm{x})}\frac{df(\bm{x})}{d\bm{x}} ,\label{l_x}
\end{align}
where $f(\bm{x})$ is the density of $\bm{X}$ at $\bm{x}$. This result is well studied and is used to obtain WADE estimators where $f(\bm{x})$, and hence $\bm{l}(\bm{x})$, is estimated through kernel methods, and no outcome regression model is used \citep{Hardle1989,Newey1994_mcfaden,Cattaneo2010}. 
Such estimators, however, are sensitive to the choice of bandwidth, and the usual asymptotic linearity of the estimator breaks down when the bandwidth is too small \citep{Cattaneo2013}. Moreover, for the ADE, i.e. when $w(\bm{x}) = 1$, the RR contains an inverse density term which can lead to extreme RR estimates that are overly sensitive to errors in the kernel density estimator around small values.

When plug-in estimators based on the derivative of an outcome regression function are used, the plug-in bias 
$\E[\bm{l}(\bm{X})\{Y - \hat{\mu}(\bm{X})\}]$ also depends on the RR, and the aforementioned issues with kernel estimation of the RR apply to plug-in bias correction procedures. Several strategies have been proposed to address these issues. \cite{Hirshberg2017, Hirshberg2018} estimate the RR using a minimax estimator that minimises the worst case mean squared error over possible regression errors in a given class. \cite{Chernozhukov2020} use a similar adversarial learning strategy to learn the RR of linear maps of regression functions, though do not apply their procedure to WADE estimation. \cite{Chernozhukov2021, Chernozhukov2022} propose a so-called automatic debiased estimator of the RR that minimises a custom loss function that is equivalent to the mean squared error in $\bm{l}(\bm{X})$. \cite{Klyne2023} estimate the RR under an assumed location-scale model for the exposure, delivering sub-optimal plug-in bias corrections when the exposure model does not hold.

Faced with difficulties in obtaining the RR of the ADE, WADEs using density weights have been proposed that facilitate estimation by removing inverse density weighting in the RR \citep{Powell1989,Cattaneo2010}. Similarly, \cite{Crump2006, Crump2009} propose weighted average treatment effect (WATE) estimands that are easier to estimate than average treatment effect (ATE) estimands because they remove inverse probability weights present in the RR of the ATE. Formally, such WATE weighting is optimal in the sense of minimising the nonparametric efficiency bound of an efficient estimator with respect to a sample analogue of the target WATE. In our work we generalise this approach to continuous exposures, as outlined in the contributions section below. Whilst optimally weighted WADEs/ WATEs might be criticised for `moving the goalposts', we believe that such estimands are appealing in settings where statistical power is restricted (e.g. due to small sample sizes or low signal to noise ratios in the true data generating mechanism), or in exploratory analyses where no specific intervention is planned.

Rather than focusing on inference of the WADE vector $\bm{\theta}$, we consider optimal weighting strategies to infer a single component, $\theta\equiv\theta_j$.
This problem is of particular interest since many practical analyses are interested in the main effect of a single continuous exposure, $A\equiv X_j$, (e.g. dose, duration, frequency), whilst accounting for other covariates (i.e. excluding the $j$th) $\bm{Z}\equiv \bm{X}_{-j}$, which may or may not be continuous.

\subsection{Contributions}

The main contribution of the current paper is to introduce and study the class of estimands $\theta = \E\{\alpha(A, \bm{Z}) Y\}$ where $\alpha$ is a member of the set
\begin{align}
    \mathcal{R} = \left\{\alpha \in \mathcal{H} \Big|
        \E\{\alpha(A,\bm{Z})A\} = 1 ,
        \E\{\alpha(A,\bm{Z})|\bm{Z}\} = 0
    \right\}. \label{riesz_hilbert}
\end{align}
and $\mathcal{H}$ is a Hilbert space, which we will define in the next section. Several example estimands of this class are presented in Section \ref{sect:riesz}. We argue that these estimands are scientifically interesting since normalised WADEs (and normalised WATEs) belong to this class, and in Section \ref{sect:riesz}, we show that each estimand in our class is a WADE (or WATE) under limited regularity assumptions. The implication of this isomorphism result is that estimands of our class can be ascribed a causal interpretation in terms of WADEs (or WATEs), with WADEs identifying so-called incremental treatment effect estimands under assumptions which we outline in Section \ref{sect:causal} \citep{Rothenhausler2019}.

Furthermore, since our class of estimands represents a unified view of WADEs and WATEs, it enables us to extend WATE results from the binary exposure setting, to new WADE results in the continuous exposure setting. In particular, we derive the estimand in our class which is optimally efficient, in the sense of minimising the efficiency bound of a sample analogue of $\theta$. This is exactly the definition of efficiency considered by \cite{Crump2006, Crump2009}, and our optimal estimand reduces to their optimal WATE estimand when the exposure is binary. When the exposure is continuous, however, then our estimand is a new optimally efficient WADE.

In the second half of our paper (Section \ref{sect_estimation}) we focus on debiased estimators for two estimands of our class, which we call `least squares estimands', due to their connections with nonparametric model projections (see discussion in Section \ref{sect:projection}),
\begin{align}
\psi &= \E\left\{ \frac{\Cov(A,Y|\bm{Z})}{\Var(A|\bm{Z})} \right\} \label{lam}
\end{align}
and
\begin{align}
\Psi &=  \frac{\E\left\{ \Cov(A,Y|\bm{Z}) \right\}}{\E\left\{ \Var(A|\bm{Z}) \right\}}.  \label{lam_bar}
\end{align}
The latter is equivalent to our optimally efficient estimand when the outcome variance is constant given exposure and covariates. Both estimands have also been studied in other contexts: $\Psi$ appears in the literature on partially-linear model estimators \citep{Vansteelandt2020,Chernozhukov2018, Newey2018}; the numerator of $\Psi$ is used for conditional independence testing \citep{Shah2018}; $\psi$ has been used to estimate the ADE under conditionally linear modelling assumptions \citep{Hirshberg2018}; when the exposure is binary, $\psi$ and $\Psi$ respectively identify the ATE and the propensity overlap weighted effect of exposure on outcome, when $\bm{Z}$ is sufficient to adjust for confounding \citep{Crump2006,Crump2009,Robins2008,Li2018,Kallus2020}. 

We compare estimators of $\psi$ and $\Psi$, the former being a contribution of our work, and the latter following from existing results. These estimators do not require estimation of the exposure density, thus alleviating the aforementioned concerns regarding kernel estimation in other WADEs ($\psi$ and $\Psi$ are WADEs according to Section \ref{sect:riesz}). Inspired by the binary exposure setting, our preferred estimator of $\psi$, is based on the R-learner of the conditional ATE \citep{Nie2017,Robinson1988}, and an analogous learner of the function $1/\Var(A|\bm{Z}=\bm{z})$, which we have not seen used elsewhere. Generally our estimators are amenable to data adaptive/ machine learning of requisite statistical functionals, as we demonstrate on simulated data in Section \ref{sect_simstud}, and on clinical data to determine the effect of Warfarin dose on blood clotting function in Section \ref{sect_appliedIll}.

\section{Weighted average derivative effects}
\label{sect:wade}

\subsection{Preliminaries}

Suppose we have $n$ iid observations, $(\bm{o}_1,...,\bm{o}_n)$ of a random variable $\bm{O}=(Y,A,\bm{Z})$ distributed according to an unknown distribution $P$, where $Y\in \mathbb{R}$ is an outcome, $A\in \mathbb{R}$ is a continuous covariate of interest which we call an `exposure' and $\bm{Z} \in \mathbb{R}^p$ is a $p$-dimensional vector of covariates. Let $\pi(\bm{Z}) \equiv \E(A|\bm{Z})$, $\mu(A,\bm{Z})\equiv\E(Y|A,\bm{Z})$ and define the WADE, $\theta_w \equiv \E\{w(A,\bm{Z})\mu^\prime(A,\bm{Z})\}$, where superscript prime denotes the derivative w.r.t. $A$, and $w(A,\bm{Z})$ is a weight such that $k\equiv\E\{w(A,\bm{Z})\}$ is finite and non-zero. We say that the weight is `normalised' when $k=1$. For the purposes of interpretation it is often desirable for the weight to be non-negative, though for full generality, we do not impose this as a restriction.
Finally, let $\mathcal{H}$ be a Hilbert space of functions $f:\mathbb{R}^{p+1} \mapsto \mathbb{R}$ equipped with inner-product $\langle f , g \rangle \equiv \E\{f(A,\bm{Z})g(A,\bm{Z})\}$ and norm $||f|| = \langle f, f \rangle^{1/2}$. We assume that $\mu \in \mathcal{H}$. Since $\mu\mapsto\theta_w$ is a bounded continuous linear map, there exists a unique RR $\alpha_\omega\in\mathcal{H}$ such that $\theta_w = \langle \mu, \alpha_\omega \rangle$, and hence $\theta_w = E\{\alpha_w(A,\bm{Z})Y\}$.

Invoking regularity conditions, \cite{Powell1989} used integration by parts (see \myappendix \ref{append:integration_by_parts}) to derive the RR of the WADE. These conditions require that $A$ is a continuous random variable and thus has a conditional density function, $f(a|\bm{z})$, given $\bm{Z}=\bm{z}$. They also require (C1) that the derivative of $w(a,\bm{z})f(a|\bm{z})$ w.r.t. $a$ exists, (C2) that $w(a,\bm{z})f(a|\bm{z})=0$ for $a$ on the boundary of the support of $A$, and (C3) that $f(a|\bm{z})= 0$ implies $w(a,\bm{z})=0$. Under these conditions
\begin{align}
\alpha_w(a,\bm{z}) &= -w^\prime(a,\bm{z}) - w(a,\bm{z}) \frac{f^\prime(a|\bm{z})}{f(a|\bm{z})}. \label{canonical_contrast}
\end{align}
Note that $\alpha_w$ refers to a single component of the RR vector in \eqref{l_x}.

\begin{remark}
    In the setting where $A\in \{0,1\}$ is a binary exposure, then one obtains the analogous Riesz representation of the WATE as $\E[w(\bm{z})\{\mu(1,\bm{z})-\mu(0,\bm{z})\}] = \langle \mu, \alpha \rangle$, where $\alpha(a,\bm{z}) = w(\bm{z})\{a-\pi(\bm{z})\}/ [\pi(\bm{z})\{1-\pi(\bm{z})\}]$ is a RR. WATEs are said to be normalised when $\E\{w(\bm{Z})\}=1$.
\end{remark}

\subsection{A Riesz representer class}
\label{sect:riesz}

Here we consider the class of estimands $\theta = \langle \mu, \alpha \rangle$ where $\alpha$ is a member of the set $\mathcal{R}$ in \eqref{riesz_hilbert}. This class is motivated by the fact that, when $k = 1$, then the RR $\alpha_w$ in \eqref{canonical_contrast} is a member of $\mathcal{R}$, i.e. normalised WADEs belong to our class of estimands. Moreover, when $A \in \{0,1\}$ is a binary exposure, then the RR of the normalised WATEs also is a member of $\mathcal{R}$, suggesting that a unified understanding of WATEs and WADEs may be obtained by studying $\mathcal{R}$. In the current paper we mostly focus on the continuous exposure setting with some additional remarks to highlight analogous results for binary exposures.
Through Theorem \ref{alpha_theorem}, we show that for each $\alpha \in \mathcal{R}$, one can construct a weight $w$ such that $\theta_w = \langle \mu, \alpha \rangle$. This implies an isomporphism between WADEs and estimands of our class. There is no guarantee, however, that the weight implied in Theorem \ref{alpha_theorem} is non-negative. We address this by deriving a sufficiency condition for weight non-negativity, which we outline in Lemma \ref{sufficiency}. The significance of these results is that, under suitable identification assumptions, then any estimand of the form $\langle \mu, \alpha \rangle$ can be ascribed a causal interpretation in terms of WADEs, see causal inference discussion in Section \ref{sect:causal}. Moreover, these results allow new WADEs to be specified by their RRs rather than their weight functions, a fact that we exploit with reference to the optimal weighting strategies in Section \ref{crump_type_result}. 

\begin{theorem}
\label{alpha_theorem}
Let $F(a|\bm{z})$ be the distribution function of $A$ given $\bm{Z}=\bm{z}$ and assume that $f(a|\bm{z})>0$ for $a$ on the convex support of $A$. For $\alpha \in \mathcal{R}$ define the weight
\begin{align}
 w(a,\bm{z}) = \frac{F(a|\bm{z})\{1-F(a|\bm{z})\}}{f(a|\bm{z})}\left[\E\{\alpha(A,\bm{Z})|A > a, \bm{Z}=\bm{z}\} - \E\{\alpha(A,\bm{Z})|A \leq a, \bm{Z}=\bm{z}\}\right]. \label{f_tilde_theorem_alpha}
\end{align}
For all differentiable functions $h \in \mathcal{H}$, $\langle h, \alpha \rangle = \E\{w(A,\bm{Z})h^\prime(A,\bm{Z})\}$ almost surely.
Proof in \myappendix \ref{main_theorem_proof}. Essentially this result is derived by treating \eqref{canonical_contrast} as an ordinary differential equation and solving for $w(a,\bm{z})$.
\end{theorem}

\begin{lemma}
\label{sufficiency}
If $\alpha(a,\bm{z}) \in \mathcal{R}$ is monotonically increasing in $a$ then the weight implied by \eqref{f_tilde_theorem_alpha} is non-negative.
Proof in \myappendix \ref{sufficiency_proof}.
\end{lemma}

\begin{remark}
\label{alpha_remark}
In the setting where $A\in \{0,1\}$ is a binary exposure, then an analogue of Theorem \eqref{alpha_theorem} is obtained by letting $w(\bm{z}) = \E\{\alpha(A,\bm{Z})A|\bm{Z}=\bm{z}\}$, with $\langle h, \alpha \rangle = \E[w(\bm{Z})\{h(1,\bm{Z}) - h(0,\bm{Z})\}]$. Thus, any estimand of the form $\langle \mu, \alpha \rangle$ can be ascribed a causal interpretation in terms of WATEs when the exposure is binary.
\end{remark}

We apply Theorem \ref{alpha_theorem} and Lemma \ref{sufficiency} in the following examples, which illustrate the connection between the RR and the WADE weight. To re-iterate, each of these examples identifies a causal effect of exposure on outcome, with weighting determined by the WADE weight.

\begin{example}[Average derivative effect (ADE)]
\label{ade}
The ADE with $w(a,\bm{z}) = 1$ was originally proposed by \cite{Hardle1989}. This results in the ADE, $\E\{\mu^\prime(A,\bm{Z})\}$ with RR $\alpha(a,\bm{z}) = -f^\prime(a|\bm{z}) / f(a|\bm{z})$, which follows immediately from \eqref{canonical_contrast}. The ADE is normalised since $\E\{w(A,\bm{Z})\}=1$ and hence $\alpha \in \mathcal{R}$. Uses of ADEs in causal inference are discussed in Section \ref{sect:causal} \citep{Rothenhausler2019}.
\end{example}

\begin{example}[Density weighted ADE]
\label{dens_ade}
Density weights are designed to facilitate inference by avoiding the inverse density weighting in the RR in Example 1 \citep{Powell1989,Cattaneo2010}.
The density weight sets $w(a,\bm{z})=f(a,\bm{z})$ with RR $- 2 f^\prime(a,\bm{z})$. This estimand is normalised to $k=\E\{f(A,\bm{Z})\}$, hence the normalised density weighted ADE is obtained by letting $w(a,\bm{z})= f(a,\bm{z})/\E\{f(A,\bm{Z})\}$ in which case the RR becomes $\alpha(a,\bm{z}) = - 2 f^\prime(a,\bm{z}) / \E\{f(A,\bm{Z})\}$ with $\alpha \in \mathcal{R}$.
\end{example}

\begin{example}[Average dose-response derivative]
\label{adrd}
The dose-response curve is a causal quantity where an intervention assigns the same exposure level for all treatment units \citep{Robins2001,Kennedy2017, Hudson2023}. Under standard assumptions, it is identified by the curve $a \mapsto \varphi(a) \equiv\E\{\mu(a,\bm{Z})\}$. Letting $k$ be a constant of normalisation, the mean derivative, $\E\{\varphi^\prime(A)\}/k$ is a WADE with weight $w(a,\bm{z})=f(a)/\{f(a|\bm{z})k\}$, where $f(a)$ is the marginal exposure density, and we make a positivity assumption such that $f(a) \neq 0 \implies f(a|\bm{z})\neq 0 $ for all $\bm{z}$. This estimand has the RR $\alpha(a,\bm{z}) = - f^\prime(a) / \{f(a|\bm{z})k\}$. When $k=\E\{f(A)/f(A|\bm{Z})\}$ then $\E\{\alpha(A,\bm{Z})A\} = \E\{w(A,\bm{Z})\} = 1$ and hence $\alpha \in \mathcal{R}$.
\end{example}

\begin{example}[Least Squares Estimands]
\label{alse}
The estimands $\psi$ and $\Psi$ in \eqref{lam} and \eqref{lam_bar} are both of the form $\langle \mu, \alpha \rangle$ with RRs
\begin{align*}
\alpha_\psi(a, \bm{z}) &\equiv  \frac{a-\E(A|\bm{Z}=\bm{z})}{\Var(A|\bm{Z}=\bm{z})} \\
\alpha_\Psi(a, \bm{z}) &\equiv  \frac{a-\E(A|\bm{Z}=\bm{z})}{\E\{\Var(A|\bm{Z}) \}}.
\end{align*}
and $\alpha_\psi, \alpha_\Psi \in \mathcal{R}$. By Theorem \ref{alpha_theorem}, this means that $\psi$ and $\Psi$ are WADEs. Applying \eqref{f_tilde_theorem_alpha} we obtain the weights
\begin{align}
w_\psi(a,\bm{z}) &= \frac{F(a|\bm{z})\{1-F(a|\bm{z})\}}{f(a|\bm{z})\Var(A|\bm{Z}=\bm{z})} \{E(A|A>a,\bm{Z}=\bm{z}) - E(A|A\leq a,\bm{Z}=\bm{z}) \}  \label{ALSE-weight-new}
\end{align}
and $w_\Psi(a,\bm{z}) = w_\psi(a,\bm{z})\Var(A|\bm{Z}=\bm{z}) / \E\{\Var(A|\bm{Z})$, respectively. Both weights are non-negative by Lemma \ref{sufficiency}. We call $\psi$ and $\Psi$ least squares estimands due to their connections with the model projections described in Section \ref{related_lit}. In \myappendix \ref{append_ls_weight} we examine in detail how $w_\psi(a,\bm{z})$ looks for various parametric exposure distributions. There is, however, no need to characterise and estimate the exposure weight to use these estimands in practice.
\end{example}

\begin{example}[Dichotomised exposure]
\label{dco_estimand}
One method for quantifying the causal effect of a continuous exposure on an outcome is via the ATE of the dichotomised exposure $\mathbb{I}(A > a_0)$ which, for a predetermined constant $a_0$, takes the value 1 when $A > a_0$ and 0 otherwise. This approach is common but discouraged on the grounds of bias, efficiency, and because it is not clear what hypothetical intervention is being considered \citep{Vanderweele2011, Berzuini2013}. Under standard assumptions for binary treatments,  however, the resulting ATE is identified by
\begin{align*}
    \E\{\E(Y|A>a_0, \bm{Z}) - \E(Y|A\leq a_0, \bm{Z})\}/k,
\end{align*}
where we introduce a normalisation constant $k$. This estimand is of the form $\langle \mu, \alpha \rangle$ with RR and implied WADE weight
\begin{align*}
  \alpha(a,\bm{z}) &= \frac{\mathbb{I}(a > a_0) - \{1 - F(a_0|\bm{z})\} }{F(a_0|\bm{z})\{1 - F(a_0|\bm{z})\} k} \\
  w(a,\bm{z}) &= \left\{ \frac{F(a|\bm{z})\{1 - F(a|\bm{z})}{F(a_0|\bm{z})\{1 - F(a_0|\bm{z})}\right\} \frac{P(A>a_0|A>a,\bm{Z}=\bm{z}) - P(A>a_0|A\leq a,\bm{Z}=\bm{z})}{f(a|\bm{z}) k}.
\end{align*}
When $k = \E\{\E(A|A>a_0, \bm{Z}) - \E(A|A\leq a_0, \bm{Z})\}$ then $\E\{\alpha(A,\bm{Z})A\} = 1$ and hence $\alpha \in \mathcal{R}$.

\end{example}

\subsection{Optimally efficient estimands}
\label{crump_type_result}

Here we derive the RR $\alpha \in \mathcal{R}$ that optimises the efficiency bound of a sample analogue of $\theta_w$. By Theorem \ref{alpha_theorem}, the optimal RR implies a corresponding optimally weighted WADE. Our chosen optimality criteria is exactly that of \cite{Crump2006,Crump2009} who derive optimal WATE weights when the exposure is binary and the weight is known (see Remark \ref{crump_remark} at the end of this Section for details). 
Thus, our contribution is to extend their method to continuous exposures with the extra subtlety being that the WADE weight depends on the exposure as well as covariates.

We rely on influence curves (ICs) to characterize the sensitivity of WADEs to small changes in the data distribution. ICs are model-free, mean zero, functionals of the true data distribution, derived from the definition of the target estimand. They are useful for constructing efficient estimators and for understanding their asymptotic efficiency bounds (see \cite{Hines2021,Fisher2020} for an introduction). This efficiency bound is a property of the estimand itself and is given by the variance of the IC, which is finite.
According to \cite{Newey1993}, when the weight function, $w(a,\bm{z})$, is known and (C1), (C2) and (C3) are assumed, the IC of $\theta_w$ is
\begin{align}
\phi_{\theta,w}(\bm{o}) =  \alpha_w(a,\bm{z}) \{y-\mu(a,\bm{z})\} + w(a,\bm{z})\mu^\prime(a,\bm{z}) -  \theta_w  \label{newey_IC}
\end{align}
where $\alpha_w(a,\bm{z})$ is the RR in \eqref{canonical_contrast} and $\bm{o}=(y,a,\bm{z})$. In all but Example \ref{ade} the weight function is unknown, however, the IC above, where the weight is known, offers some insight into optimal weight selection. Specifically, we minimize the efficiency bound of an efficient estimator, $\hat{\theta}_w$, of the sample analogue of $\theta_w$,
\begin{align*}
\theta_{w,S} &\equiv n^{-1}\sum_{i=1}^n w(a_i,\bm{z}_i)\mu^\prime(a_i,\bm{z}_i) \\
\sqrt{n}(\hat{\theta}_w-\theta_{w,S}) & \overset{d}{\to} \n{0}{V} \\
V &\equiv \E\{\alpha_w^2(A,\bm{Z})\sigma^2(A,\bm{Z})\}
\end{align*}
where $\sigma^2(A,\bm{Z})\equiv \Var(Y|A,\bm{Z})$. The efficiency bound with respect to $\theta_{w,S}$, rather than $\theta_w$, is chosen so that the final two terms in \eqref{newey_IC} may be disregarded. Not only does this simplify the subsequent analysis, but these terms capture the difference between the WADE conditional on the sample distribution and that of the population as a whole, which depends on the unknown value of $\theta_w$. I.e.,
\begin{align*}
\sqrt{n}(\hat{\theta}_w-\theta_{w}) & \overset{d}{\to} \n{0}{V+U} \\
U &\equiv \E[\{w(A,\bm{Z})\mu^\prime(A,\bm{Z}) -  \theta_w\}^2]
\end{align*}
thus selecting weights to minimise $V+U$ is conceptually problematic as $\theta_w$ is itself the target estimand \citep{Crump2006,Crump2009}. Theorem \ref{alpha_theorem}, offers constraints on the contrast function under which $\alpha_w(a,\bm{z})$ is the RR of a weighted ADE. Our goal, therefore, is to find $\alpha_w \in \mathcal{R}$ which minimises $V$. The optimal solution is given in general by Theorem \ref{optimality_1}, with Lemma \ref{sufficiency} ensuring that the resulting RR is non-negative.

\begin{theorem}
\label{optimality_1}
Minimizing the efficiency bound $V=n\Var\{\hat{\theta}_w-\theta_{w,S}\}$, subject to the constraints, $\E\{\alpha_w(A,\bm{Z})|\bm{Z}\}=0$, and $\E\{\alpha_w(A,\bm{Z})\}=1$, has the solution
\begin{align}
  \alpha_w(a,\bm{z}) &= 
  \frac{\{a - \tilde{\pi}(\bm{z})\} / \sigma^2(a, \bm{z}) }{ \E[A\{A - \tilde{\pi}(\bm{Z})\}/ \sigma^2(A, \bm{Z}) ] } , \label{optimal_rr}\\
  \tilde{\pi}(\bm{z}) &\equiv \frac{\E\{A / \sigma^2(A,\bm{Z}) | \bm{Z} = \bm{z}\}}{\E\{1 / \sigma^2(A,\bm{Z}) | \bm{Z} = \bm{z}\}} \nonumber
\end{align}
which implies the optimally efficient estimand
\begin{align*}
  \frac{\E\left[Y\{A - \tilde{\pi}(\bm{Z})\}/ \sigma^2(A, \bm{Z})\right]}{\E\left[A \{A - \tilde{\pi}(\bm{Z})\} / \sigma^2(A, \bm{Z})\right]}.
\end{align*}
Proof in \myappendix \ref{optimality_proof}.
\end{theorem}
\begin{corollary}
\label{indep_corol}
When $Y$ is homoscedastic conditional on $\bm{Z}$, i.e. $\sigma^2(a,\bm{z})=\sigma^2(\bm{z}) \equiv \Var(Y|\bm{Z}=\bm{z})$ then the estimand in Theorem \ref{optimality_1} is
\begin{align*}
\frac{\E\{\Cov(A,Y|\bm{Z})/\sigma^2(\bm{Z})\}}{\E\{\Var(A|\bm{Z})/\sigma^2(\bm{Z})\}}.
\end{align*}
For proof, observe that under conditional homoscedasticity, $\tilde{\pi}(\bm{z}) = \pi(\bm{z})$. Furthermore, when $Y$ is homoscedastic, i.e. $\sigma^2(a,\bm{z})$ is constant, then the optimal estimand is $\Psi$.
\end{corollary}

\begin{remark}
\label{crump_remark}

In the setting where $A\in \{0,1\}$ is a binary exposure, \cite{Crump2006,Crump2009} consider the WATE $\E[w(\bm{Z}) \{\mu(1,\bm{Z}) - \mu(0,\bm{Z})\}]$ where $w(\bm{z})$ is known. They derive that the efficiency bound of a sample analogue  WATE has the same form as $V$ above, with RR $\alpha(a,\bm{z}) = w(\bm{z})\{a-\pi(\bm{Z})\}/[\pi(\bm{z})\{1-\pi(\bm{z})\}]$. They minimise $V$ over $w(\bm{z})$ with the normalisation constraint $\E\{w(\bm{Z})\} = 1$, and their result is recovered by Theorem \ref{optimality_1} when $A$ is binary (see \myappendix \ref{crump_appendix} for details). Thus, Theorem \ref{optimality_1} represents a generalisation that covers continuous and binary exposures.
\end{remark}

\section{Estimation}
\label{sect_estimation}
\subsection{Efficient estimators}

Here we focus on efficient estimation of $\psi$ and $\Psi$ as in \eqref{lam} and \eqref{lam_bar}. The ICs of $\psi$ and $\Psi$ respectively are,
\begin{align*}
\phi_{\psi}(\bm{o}) &= \frac{\{a-\pi(\bm{z})\}}{\beta(\bm{z})}\left[y - \mu(\bm{z}) - \lambda(\bm{z})\{a-\pi(\bm{z})\} \right] +  \lambda(\bm{z})- \psi \\ 
\phi_{\Psi}(\bm{o}) &= \frac{\{a-\pi(\bm{z})\}}{\E\{\beta(\bm{Z})\}}\left[y - \mu(\bm{z}) - \Psi \{a-\pi(\bm{z})\}\right],  
\end{align*}
where $\beta(\bm{z})\equiv \Var(A|\bm{Z}=\bm{z})$. These ICs may be used to construct efficient estimating equation estimators of $\psi$ and $\Psi$ by setting (an estimate of) the sample mean IC to zero. For $\psi$, this strategy is equivalent to the so-called one-step correction which we outline in \myappendix \ref{appendix:asymptotics}. For $\psi$ and $\Psi$, we thus obtain the estimators
\begin{align*}
\hat{\psi} &= n^{-1}\sum_{i=1}^n  \frac{\{a_i - \hat{\pi}(\bm{z}_i)\}}{\hat{\beta}(\bm{z}_i)}[y_i - \hat{\mu}(\bm{z}_i) - \hat{\lambda}(\bm{z}_i)\{a_i - \hat{\pi}(\bm{z}_i)\}] +  \hat{\lambda}(\bm{z}_i) \\ 
\hat{\Psi} &= \frac{\sum_{i=1}^n\{a_i - \hat{\pi}(\bm{z}_i)\}\{y_i - \hat{\mu}(\bm{z}_i)\}}{\sum_{i=1}^n\{a_i - \hat{\pi}(\bm{z}_i)\}^2},
\end{align*}
where superscript hat denotes consistent estimators. In practice, we recommend a cross-fitting approach of the type described in Section \ref{proposed_algos}, to obtain the fitted models and evaluate the estimators using a single sample \citep{Chernozhukov2018,van_der_laan_cross-validated_2011}. We discuss the reasons for sample splitting with reference to Theorems \ref{asym_theorem_alse1} and \ref{asym_theorem_alse2}, which give conditions under which $\hat{\psi}$ and $\hat{\Psi}$ are regular asymptotically linear (RAL). Note that Theorem \ref{asym_theorem_alse1} is a novel contribution of the current paper, but similar Theorems to \ref{asym_theorem_alse2} have been considered by others \citep{Chernozhukov2018, Xiang2020}.

\begin{theorem}
\label{asym_theorem_alse1}
Assume that there exists constants $\epsilon>0, K>0$ such that (almost surely) $\epsilon < \hat{\beta}(\bm{Z})$, $\beta(\bm{Z}) \in (\epsilon, K) $, $\hat{\lambda}(\bm{Z})\in(-K,K)$, $\Var(Y|\bm{Z}) < K$, $E[\{Y-\mu(\bm{Z})\}^4 |\bm{Z}] < K$, $E[\{A-\pi(\bm{Z})\}^4 |\bm{Z}] < K$. Suppose also that at least one of the following two conditions hold:
\begin{enumerate}[label=\arabic*.]
\item (Sample-splitting) $\hat{\pi}(\bm{z}), \hat{\mu}(\bm{z}), \hat{\lambda}(\bm{z})$, and $\hat{\beta}(\bm{z})$ are obtained from a sample independent of the one used to construct $\hat{\psi}$.
\item (Donsker condition) The quantities $\hat{\lambda}(\bm{Z})$,
\begin{align*}
\frac{\{A-\hat{\pi}(\bm{Z})\}\{Y - \hat{\mu}(\bm{Z})\}}{\hat{\beta}(\bm{Z})}, \frac{\hat{\lambda}(\bm{Z})\{A-\hat{\pi}(\bm{Z})\}^2}{\hat{\beta}(\bm{Z})} 
\end{align*}
fall within a $P$-Donsker class with probability approaching $1$.
\end{enumerate}
Finally, letting $||.||$ denote the $\mathcal{L}_2(\bm{Z})$ norm, assume
\begin{enumerate}[label=(A\arabic*)]
\item $||\pi-\hat{\pi}|| = o_P(n^{-\nu/4})$ and $||\mu-\hat{\mu}|| = o_P(n^{-\tau/4})$ where $\nu \geq1$, $\tau\geq0$ and $\nu + \tau \geq 2$.
\item The product of $||\lambda-\hat{\lambda}||$ and $||\beta-\hat{\beta}||$ is $o_P(n^{-1/2})$.
\end{enumerate}
Then $\hat{\psi}$ is RAL with IC, $\phi_{\psi}(\bm{\bm{O}})$, and hence $\sqrt{n}(\hat{\psi}-\psi)$ converges in distribution to a mean-zero normal random variable with variance $\E\{\phi_{\psi}^2(\bm{\bm{O}})\}$.
Proof in \myappendix \ref{appendix:asymptotics}.
\end{theorem}

\begin{theorem}
\label{asym_theorem_alse2}
Assume (A1) in Theorem \ref{asym_theorem_alse1}, the quantity $n^{-1}\sum_{i=1}^n\{a_i - \hat{\pi}(\bm{z}_i)\}^2 > 0$, and $\E[\{A-\pi(\bm{Z})\}^2] > 0$, there exists a constant $K>0$ such that $\Var(Y|\bm{Z}) < K$ and $\beta(\bm{Z}) < K$ and suppose that at least one of the following two conditions hold:
\begin{enumerate}[label=\arabic*.]
\item (Sample-splitting) $\hat{\pi}(\bm{z})$ and $ \hat{\mu}(\bm{z})$ are obtained from a sample independent of the one used to construct $\hat{\Psi}$.
\item (Donsker condition) The quantities $\{A-\hat{\pi}(\bm{Z})\}\{Y - \hat{\mu}(\bm{Z})\}$ and $\{A-\hat{\pi}(\bm{Z})\}^2$ fall within a $P$-Donsker class with probability approaching $1$.
\end{enumerate}
Then $\hat{\Psi}$ is RAL with IC, $\phi_{\Psi}(\bm{\bm{O}})$, and hence $\sqrt{n}(\hat{\Psi}-\Psi)$ converges in distribution to a mean-zero normal random variable with variance $\E\{\phi_{\Psi}^2(\bm{\bm{O}})\}$.
Proof in \myappendix \ref{appendix:asymptotics}.
\end{theorem}

The estimator $\hat{\psi}$ requires learning the functions $\beta$ and $\lambda$, whereas $\hat{\Psi}$ does not, with Theorem \ref{asym_theorem_alse1} requiring (A2) to control the error in $\hat{\beta}$ and $\hat{\lambda}$. This distinction makes $\Psi$ generally more straightforward to estimate than $\psi$. Assumption (A2) also demonstrates that $\hat{\psi}$ is `rate double robust', in the sense $\hat{\lambda}$ can converge slowly, so long as $\hat{\beta}$, converges sufficiently quickly, and vice-versa. Thus, one can trade-off accuracy in $\hat{\lambda}$ and $\hat{\beta}$. Similar double robustness has been demonstrated previously, e.g. for the augmented inverse probability weighted (AIPW) estimator of the ATE \citep{Robins1994}, which trades-off accuracy in the propensity score and outcome estimators. On top of this double robustness, (A1) implies a one-sided robustness of $\hat{\psi}$ and $\hat{\Psi}$ with respect to the estimators $\hat{\pi}$, and $\hat{\mu}$. Specifically, $\hat{\mu}$ can converge slowly, so long as $\hat{\pi}$ converges sufficiently quickly, but the converse is not true, since (A1) requires that $\hat{\pi}$ converges at least at $n^{1/4}$ rate.

The Donsker conditions in Theorems \ref{asym_theorem_alse1} and \ref{asym_theorem_alse2} are usually not guaranteed to hold when flexible machine learning methods are used to estimate nuisance functions. Fortunately, sample splitting/ cross fitting of nuisance functions offers a way of avoiding Donsker conditions, at the expense of making nuisance functions more computationally expensive to learn \citep{Chernozhukov2018,van_der_laan_cross-validated_2011}. Moreover, the estimator $\hat{\Psi}$ has been studied before in the context of partially linear models \citep{Robinson1988} and nonparametric estimation of GLM coefficients \citep{Vansteelandt2020}. Its properties are an active research topic with regards to the smoothness and convergence rates of nuisance estimators \citep{Newey2018,Balakrishnan2023}.

\begin{remark}
In a setting where $A\in\{0,1\}$ is a binary exposure then $\hat{\psi}$ reduces to the AIPW estimator of the ATE. In particular, if we estimate $\hat{\mu}(a,\bm{z}), \hat{\pi}(\bm{z})$, and let $\hat{\lambda}(\bm{z})=\hat{\mu}(1,\bm{z})-\hat{\mu}(0,\bm{z})$, $\hat{\beta}(\bm{z})=\hat{\pi}(\bm{z})\{1-\hat{\pi}(\bm{z})\}$, and $\hat{\mu}(\bm{z}) =\hat{\mu}(0,\bm{z}) + \hat{\lambda}(\bm{z})\hat{\pi}(\bm{z})$, then one obtains the AIPW estimator
\begin{align*}
n^{-1}\sum_{i=1}^n  \frac{\{a_i - \hat{\pi}(\bm{z}_i)\}}{\hat{\pi}(\bm{z}_i)\{1-\hat{\pi}(\bm{z}_i)\}}\{y_i - \hat{\mu}(a_i,\bm{z}_i)\} +  \hat{\mu}(1,\bm{z}_i)-\hat{\mu}(0,\bm{z}_i).
\end{align*}
Hence, $\hat{\psi}$ represents a generalisation of the AIPW estimator that covers binary and continuous exposures.
\end{remark}

\subsection{Nuisance function estimators}

The estimator $\hat{\Psi}$ is indexed by the choice of estimator for $\hat{\mu}$ and $\hat{\pi}$, with the estimator $\hat{\psi}$ additionally indexed by the choice of estimator for $\hat{\lambda}$ and $\hat{\beta}$. Generally, we are not constrained to any particular learning method, making these estimators amenable to data adaptive/ machine learning estimation of these working models.

Data adaptive regression algorithms are well developed for the regularised regression of an observed variable on to a set of explanatory variables, e.g. for the functions $\mu$, and $\pi$ in the present context, which can be estimated by respectively regressing $Y$ and $A$ on $\bm{Z}$. For $\lambda$ and $\beta$, however, estimation methods are less well developed, and we propose so-called meta-learning approaches, which estimate $\lambda$ and $\beta$ by solving a series of regression problems.

In the setting where $A\in\{0,1\}$ is a binary exposure, $\lambda$ represents the conditional ATE, estimation of which is a highly active area of research, with an emphasis on flexible machine learning methods \citep{Abrevaya2015,Athey2016,Nie2017,Kallus2018,Wager2018,Kunzel2019,Kennedy2020}. Also in the binary exposure setting, $\beta(\bm{z}) = \pi(\bm{z})\{1-\pi(\bm{z})\}$ hence there is no need for a separate estimator of $\beta$. The problem of estimating conditional variance functions more generally, however, has received some attention in the literature, with applications in constructing confidence intervals for the conditional mean function $\pi$ and for estimating signal-to-noise ratios \citep{Shen2020,Wang2008,Cai2009,Verzelen2018}. We will consider two approaches to estimating $\lambda$ and $\beta$.

The first approach, which we shall refer to as the direct learning approach, involves decomposing $\lambda$ and $\beta$ into functions of conditional expectations, each of which can be estimated using standard regression methods, with the estimates combined to produce $\hat{\lambda}$ and $\hat{\beta}$. Specifically, letting $\Ehat\{YA|\bm{Z}=\bm{z}\}$ and $\Ehat\{A^2|\bm{Z}=\bm{z}\}$ denote estimates obtained by respectively regressing $YA$ and $A^2$ on $\bm{Z}$, we define nuisance estimators
\begin{align}
\hat{\lambda}(\bm{z}) &= \frac{\Ehat\{YA|\bm{Z}=\bm{z}\} - \hat{\mu}(\bm{z})\hat{\pi}(\bm{z})}{\Ehat\{A^2|\bm{Z}=\bm{z}\} - \hat{\pi}^2(\bm{z})} \label{dir_lambda} \\
\hat{\beta}(\bm{z})   &= \Ehat\{A^2|\bm{Z}=\bm{z}\} - \hat{\pi}^2(\bm{z}). \label{dir_beta}
\end{align}
The issue with this direct approach, however, is that whilst regularization methods can be used to control the smoothness of each individual regression function, there is no guarantee on the smoothness of $\hat{\lambda}$ and $\hat{\beta}$. In practice these may be erratic functions due to artefacts of the regularization of the individual regression functions. Additionally, there is no guarantee that $\hat{\beta}$, which also represents the denominator of $\hat{\lambda}$, is greater than zero.

Moreover, when $\lambda$ or $\beta$ are simple functions (e.g. constant), one might hope that they would be easy to learn at fast rates of convergence. The corresponding direct estimators, however, might inherit slow convergence rates from the estimators $\hat{\mu}, \hat{\pi}, \ldots$, which could be the case when these functions are, in truth, complex. These issues motivate an alternative approach where the complexities of $\hat{\lambda}$ and $\hat{\beta}$ can be controlled directly, and one can ensure that $\hat{\beta}(\bm{z})>0$.

The second approach, which we shall refer to as the quasi-oracle learning approach, is a meta-learning method based on the R-learner of the conditional ATE \citep{Nie2017,Robinson1988}. In our description we make use of the following Lemma.

\begin{lemma}
\label{R_identLemma}
Let $\bm{O}=(\bm{Z},V,W)$ be a random variable consisting of $\bm{Z}\in\mathbb{R}^p$, $V\in\mathbb{R}$, and  $W\in\mathbb{R}$ with $W > 0$ almost surely. Let $\mathcal{F}$ denote the set of all possible functions $g:\mathbb{R}^p\mapsto \mathbb{R}$. Then
\begin{align}
\frac{\E(V|\bm{Z}=\bm{z})}{\E(W|\bm{Z}=\bm{z})} = \argmin_{g \in \mathcal{F}} \E\left[ W\left\{ \frac{V}{W} - g(\bm{Z}) \right\}^2\right]
\label{R-identity}
\end{align}
where we say the part in the square brackets is to equal 0 when $W=0$ and requisite moments of $\bm{O}$ are assumed to be finite. See \myappendix \ref{appendix:optimal_derivs} for proof.
\end{lemma}

This Lemma connects the problem of estimating a ratio of conditional expectations, with minimisation of a weighted mean squared error. For example, in the setting where $W=1$, then the right hand side of \eqref{R-identity} reduces to the familiar mean squared error loss. Similarly, the left hand side of \eqref{R-identity} recovers $\lambda(\bm{z})$ in the setting where $W=\{A-\pi(\bm{Z})\}^2$ and $V=\{Y-\mu(\bm{Z})\}\{A-\pi(\bm{Z})\}$, suggesting that an estimator for $\lambda$ is obtained by regressing $\{Y-\mu(\bm{Z})\}/\{A-\pi(\bm{Z})\}$ on $\bm{Z}$ with weights $\{A-\pi(\bm{Z})\}^2$. 

We call this an `oracle estimator' for $\lambda$, since it is the regression problem that we would like to solve if these outcomes and weights were known. The R-learner of the conditional ATE essentially mimics the oracle learner by first estimating $\mu$ and $\pi$ using an independent sample, then using these to estimate the unobserved outcomes and weights. This method is referred to as `quasi-oracle' since the error bound for the $\lambda$ estimator may decay faster than those of the $\mu$ and $\pi$ estimators \citep{Nie2017}.

We propose a similar approach to learning $\beta$, which appears in our target estimator $\hat{\psi}$ as an inverse weight. Such inverse weighting may be problematic when $\beta(\bm{z})$ is in truth small, since small errors in $\hat{\beta}(\bm{z})$ could result in large differences in the value of $1/\hat{\beta}(\bm{z})$. This extreme weighting problem is well documented in the context of inverse probability weighting estimators of the ATE \citep{Kang2007}. Concerns regarding extreme weights, however, could be mitigated by regularizing the function $1/\hat{\beta}$ rather than $\hat{\beta}$ itself. For this reason we consider that the left hand side of \eqref{R-identity} recovers $1/\beta(\bm{z})$ in the setting where $V=1$ and $W=\{A-\pi(\bm{Z})\}^2$.

This suggests that an oracle estimator for $1/\beta$ is obtained by regressing $\{A-\pi(\bm{Z})\}^{-2}$ on $\bm{Z}$ with weights $\{A-\pi(\bm{Z})\}^2$. Analogous to the the R-learner, we propose a learner which mimics this oracle learner by first estimating $\pi$ using an independent sample, then estimating the oracle outcomes and weights.

\subsection{Proposed algorithms}
\label{proposed_algos}

The proposed working function estimators are implemented in Algorithms noSS and SS below. The latter uses a cross fitting regime to ensure that $\hat{\mu}(\bm{z}_i),\hat{\pi}(\bm{z}_i),\hat{\lambda}(\bm{z}_i)$, and $\hat{\beta}(\bm{z}_i)$ are obtained using working models which are constructed from a dataset that does not include the $i$th observation. This is useful in controlling the so-called empirical process term \citep{Chernozhukov2018,van_der_laan_cross-validated_2011}.

Algorithms noSS and SS return estimates $\{\hat{\pi}_i\}_{i=1}^n$, $\{\hat{\mu}_i\}_{i=1}^n$, $\{\hat{\lambda}_i\}_{i=1}^n$, and $\{\hat{\beta}_i\}_{i=1}^n$, which can be used to obtain
\begin{align*}
\hat{\psi} &= n^{-1}\sum_{i=1}^n  \frac{(a_i - \hat{\pi}_i)}{\hat{\beta}_i}\{y_i - \hat{\mu}_i - \hat{\lambda}_i(a_i - \hat{\pi}_i)\} +  \hat{\lambda}_i\\
\hat{\Psi} &= \frac{\sum_{i=1}^n(a_i - \hat{\pi}_i)(y_i - \hat{\mu}_i)}{\sum_{i=1}^n(a_i - \hat{\pi}_i)^2},
\end{align*}
with variances respectively estimated by $n^{-2}\sum_{i=1}^n \phi_{\psi,i}^2$ and $n^{-2}\sum_{i=1}^n \phi_{\Psi,i}^2$ where
\begin{align*}
\phi_{\psi,i} &= \frac{(a_i - \hat{\pi}_i)}{\hat{\beta}_i}\{y_i - \hat{\mu}_i - \hat{\lambda}_i(a_i - \hat{\pi}_i)\} +  \hat{\lambda}_i - \hat{\psi}\\
\phi_{\Psi,i} &= \frac{(a_i - \hat{\pi}_i)}{\hat{\eta}} \{y_i - \hat{\mu}_i - \hat{\Psi}(a_i - \hat{\pi}_i)\}
\end{align*}
and $\hat{\eta} \equiv n^{-1}\sum_{i=1}^n (a_i - \hat{\pi}_i)^2$ is an estimate of $\E\{\beta(\bm{Z})\}$. We note that where the algorithms require regression estimates to be `fitted', any suitable regression/ machine learning method can be used.

Both algorithms are also indexed by the choice of learner for $\hat{\lambda}$ and $\hat{\beta}$ in steps 2 and 3 of each algorithm respectively, with the substeps marked (A) and (B) referring to the direct, and quasi-oracle approaches. Note that the quasi-oracle methods in these algorithms do not use sample splitting to learn the unobserved outcomes and weights, this is due to the impracticality of the extra sample splitting required to do so. Substeps (A) and (B) do not need to be carried out for inference of $\hat{\Psi}$ only.

For estimators such as $\hat{\Psi}$, it has been suggested that faster convergence rates may be attained through additional sample splitting, which ensures that $\hat{\mu}(\bm{z}_i)$ and $\hat{\pi}(\bm{z}_i)$ are obtained from two different and independent datasets, neither of which containing the $i$th observation \citep{Newey2018}. We do not consider such `double cross fitting' here, since extensions, to estimate $\psi$, would require significant additional sample splitting to estimate $\{\hat{\lambda}_i\}_{i=1}^n$ and $\{\hat{\beta}_i\}_{i=1}^n$,  which may be impractical in finite samples.

\textbf{Algorithm noSS - no sample splitting}
\begin{enumerate}[label=(\arabic*)]
\item Fit $\hat{\mu}(\bm{z})$ and $\hat{\pi}(\bm{z})$. Use these fitted models to obtain $\hat{\mu}_i \equiv\hat{\mu}(\bm{z}_i)$ and $\hat{\pi}_i \equiv\hat{\pi}(\bm{z}_i)$.
\item (A) Fit $\Ehat\{YA|\bm{Z}=\bm{z}\}$ and $\Ehat\{A^2|\bm{Z}=\bm{z}\}$ and use these to construct $\hat{\lambda}(\bm{z})$ and $\hat{\beta}(\bm{z})$ as in \eqref{dir_lambda} and \eqref{dir_beta}. Or (B) obtain $\hat{\lambda}(\bm{z})$ and $1/\hat{\beta}(\bm{z})$ respectively by regressing $\{Y-\hat{\mu}(\bm{Z})\}/\{A-\hat{\pi}(\bm{Z})\}$ and $\{A-\hat{\pi}(\bm{Z})\}^{-2}$ on $\bm{Z}$ with weights $\{A-\hat{\pi}(\bm{Z})\}^2$ using all the data. After doing (A) or (B), use the fitted models to obtain $\hat{\lambda}_i \equiv\hat{\lambda}(\bm{z}_i)$ and $\hat{\beta}_i \equiv\hat{\beta}(\bm{z}_i)$.
\end{enumerate}

\textbf{Algorithm SS - sample splitting}
\begin{enumerate}[label=(\arabic*)]
\item Split the data into $K$ folds.
\item \textbf{For} each fold $k$: Fit $\hat{\mu}(\bm{z})$ and $\hat{\pi}(\bm{z})$ using the data set excluding fold $k$. Use these fitted models to obtain $\hat{\mu}_i \equiv\hat{\mu}(\bm{z}_i)$ and $\hat{\pi}_i \equiv\hat{\pi}(\bm{z}_i)$ for $i$ in fold $k$.
\item (A) Fit $\Ehat\{YA|\bm{Z}=\bm{z}\}$ and $\Ehat\{A^2|\bm{Z}=\bm{z}\}$ using the data set excluding fold $k$, and use these to construct $\hat{\lambda}(\bm{z})$ and $\hat{\beta}(\bm{z})$ as in \eqref{dir_lambda} and \eqref{dir_beta}. Or (B) obtain $\hat{\lambda}(\bm{z})$ and $1/\hat{\beta}(\bm{z})$ respectively by regressing $\{Y-\hat{\mu}(\bm{Z})\}/\{A-\hat{\pi}(\bm{Z})\}$ and $\{A-\hat{\pi}(\bm{Z})\}^{-2}$ on $\bm{Z}$ with weights $\{A-\hat{\pi}(\bm{Z})\}^2$ using the data set excluding fold $k$. After doing (A) or (B), use the fitted models to obtain $\hat{\lambda}_i \equiv\hat{\lambda}(\bm{z}_i)$ and $\hat{\beta}_i \equiv\hat{\beta}(\bm{z}_i)$ for $i$ in fold $k$. \textbf{End for.}
\end{enumerate}

\section{Simulation study}
\label{sect_simstud}

In our simulation study we compared Algorithms noSS and SS for estimating $\Psi$ and Algorithms noSS-A, noSS-B, SS-A and SS-B for estimating $\psi$ on generated data in finite samples, using $K=5$ fold sample splitting. Reproduction code for this study is available at \verb|github.com/ohines/alse|. We generated $1000$ datasets of size $n\in \{500,1000,...,4000\}$ from the following structural equation model
\begin{align*}
Z_1,Z_2,Z_3 &\sim \text{Uniform}(-1,1) \\
\epsilon_1,\epsilon_2 &\sim \n{0}{1}\\
A &= Z_1+0.5Z_1^3-2Z_2^2 + Z_1^2Z_2 + (1+Z_1^2)\epsilon_1\\
Y &= A(1+Z_1-Z_1^2-0.5Z_2^2) -Z_1^2Z_2 + Z_2Z_3  + \epsilon_2
\end{align*}
with the least squares estimands taking the true values $\psi = 0.5$ and $\Psi =107/294 \approx 0.36$.

For each dataset, $\hat{\psi}$ and $\hat{\Psi}$ were estimated along with their variance and associated Wald based (95\%) confidence intervals. Two regression model approaches were considered, the first used generalised additive models, as implemented through the \verb|mgcv| package in R \citep{Wood2016}. These models use flexible spline smoothing including pairwise interaction terms. The second regression modelling approach used random forest learners available through the \verb|ranger| package in R \citep{Wright2017}.

\begin{figure}[htbp]
\centering
\includegraphics[width=\linewidth]{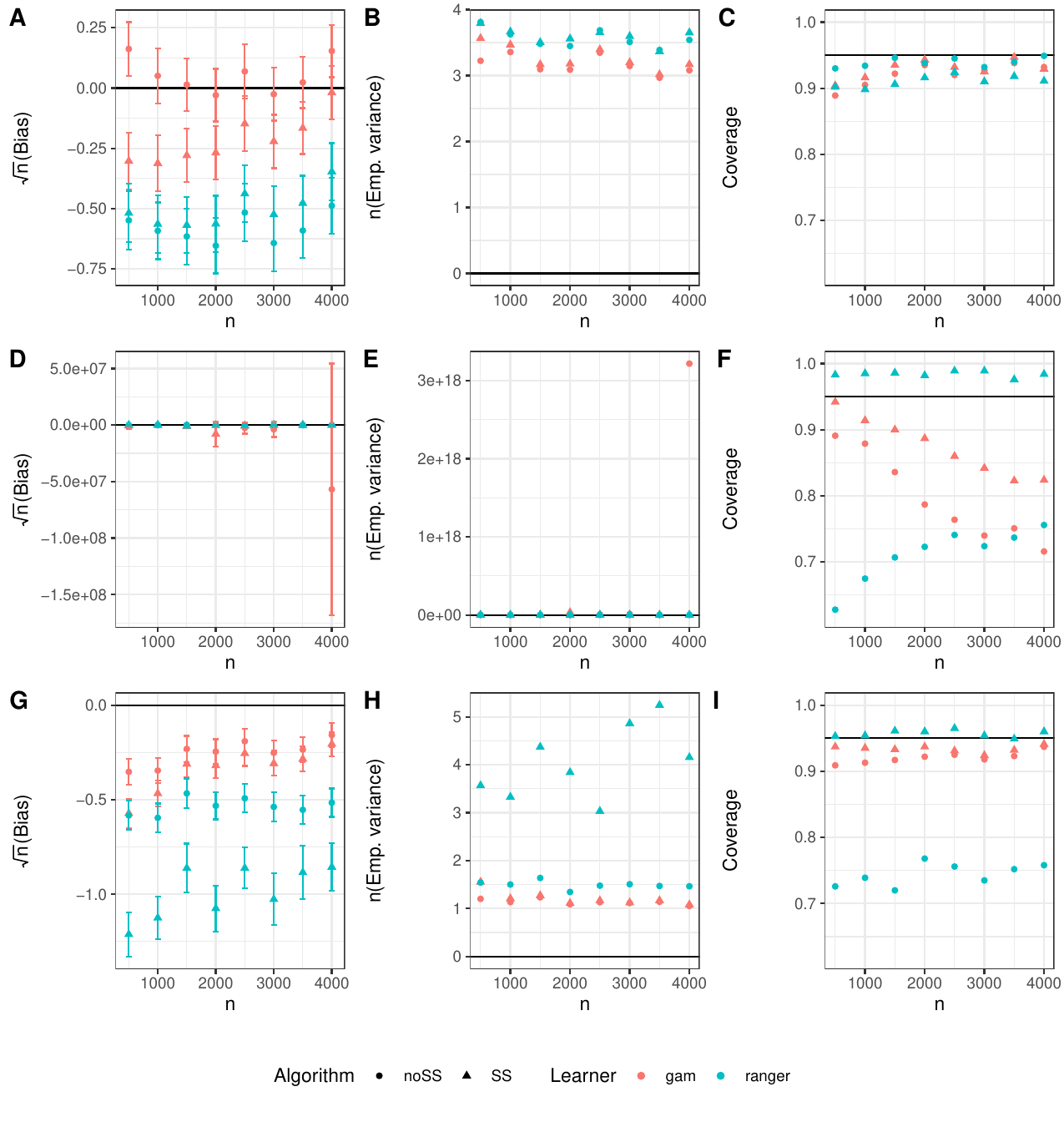}
\caption{Sample size against Bias (1st column), variance (2nd) and 95\% Wald CI coverage (3rd) for $\hat{\Psi}$ (plots A,B,C), $\hat{\psi}$ using the direct approach (plots D,E,F), and $\hat{\psi}$ using the quasi-oracle approach (plots G,H,I). We highlight that the y-axis limits change between row of the bias and variance plots. The black horizontal line represents 0 bias and variance, and 95\% coverage respectively.
}
\label{sim_plot}
\end{figure}

Figure \ref{sim_plot} shows empirical estimates of the empirical bias and empirical variance of $\hat{\psi}$ and $\hat{\Psi}$ scaled by $\sqrt{n}$ and $n$ respectively, as well as the empirical coverage probability of a Wald based 95\% confidence-interval. Comparing Algorithms noSS and SS for the estimation of $\hat{\psi}$, we notice that sample splitting generally improves confidence interval coverage.

Additionally, for estimation of $\psi$ the quasi-oracle approach (Algorithm -B) outperforms the direct approach (Algorithm -A) in terms of reduced bias, variance and improved CI coverage. This is achieved since the quasi-oracle approach controls the smoothness of $\hat{\lambda}$ and the inverse weights $1/\hat{\beta}$, whereas Algorithm -A does not, leading to the possibility of extreme inverse weighting in the estimator. Indeed, the extreme bias values presented for Algorithm -A are due to extreme negative estimates of $1/\beta(\bm{Z})$, which are corrected by the quasi-oracle estimator (Algorithm -B). On the basis of these results, we recommend Algorithm noSS-B for estimation of $\psi$ and Algorithm noSS for estimation of $\Psi$.

\section{Warfarin dose example}
\label{sect_appliedIll}

We illustrate the proposed estimators using the IWPC \citep{Cunningham2009} dataset, which has also been reanalysed several times in literature on dynamic treatment rule estimation \citep{Schulz2021,Wallace2018,Chen2016}. Reproduction code for this analysis is available at \verb|github.com/ohines/alse|. The data consists of $n=1732$ patients receiving Warfarin therapy, which is a commonly prescribed  anticoagulant used to treat thrombosis and thromboembolism. We consider least squares estimands for the effect of Warfarin dose $(A)$ on international normalised ratio (INR)  $(Y)$, which is a measure of blood clotting function, given 13 other patient characteristics $(\bm{Z})$, including genetic data, as described in \cite{Cunningham2009}.

Fitted models were obtained using the Super Learner \citep{VanDerLaan2007}, an ensemble learning method, implemented in the \verb|SuperLearner| package in R. This used 20 cross validation folds, and a `learner library' containing various routines (\verb|glm|, \verb|glmnet|, \verb|gam|, \verb|xgboost|, \verb|ranger|). Additional results which use the `discrete' Super Learner for model fitting are presented in \myappendix \ref{appendix:iwpc_extra}. The discrete Super Learner selects the regression algorithm in the learner library which minimises a cross validated estimate of e.g. the mean squared error loss, whereas the Super Learner minimizes the same loss by taking a convex combination of learners. For the sample splitting algorithms (Algorithm SS), $K=20$ folds were chosen (between 10 to 20 folds is typical for cross-fitting procedures).

The results, presented in Table \ref{IWPCresults}, suggest that increased Warfarin is associated with an increase in INR. We see that the estimators for $\Psi$ tend to have narrower confidence intervals than estimators for $\psi$, with commensurately smaller Wald based P-values for the corresponding `zero-effect' null (i.e. the null hypothesis that $\Psi = 0$ vs. the null hypothesis that $\psi=0$). This type of difference is to be expected, based on the efficiency arguments presented in Section \ref{crump_type_result}. Additionally, the estimators for $\psi$, which use the R-learner for conditional effect estimation (Algorithms noSS-B and SS-B) give more credible estimates than those that use the direct approach (Algorithms noSS-A and SS-A), in the sense that they are of a similar order of magnitude to the $\Psi$ estimates. Moreover, we see that sample splitting leads to more credible estimates, compared with no sample splitting, as evident in Algorithms SS-A versus noSS-A. This difference is because sample splitting helps to control for overfitting of the functional estimators.

\begin{figure}[htb]
    \centering
    \includegraphics[width=\linewidth]{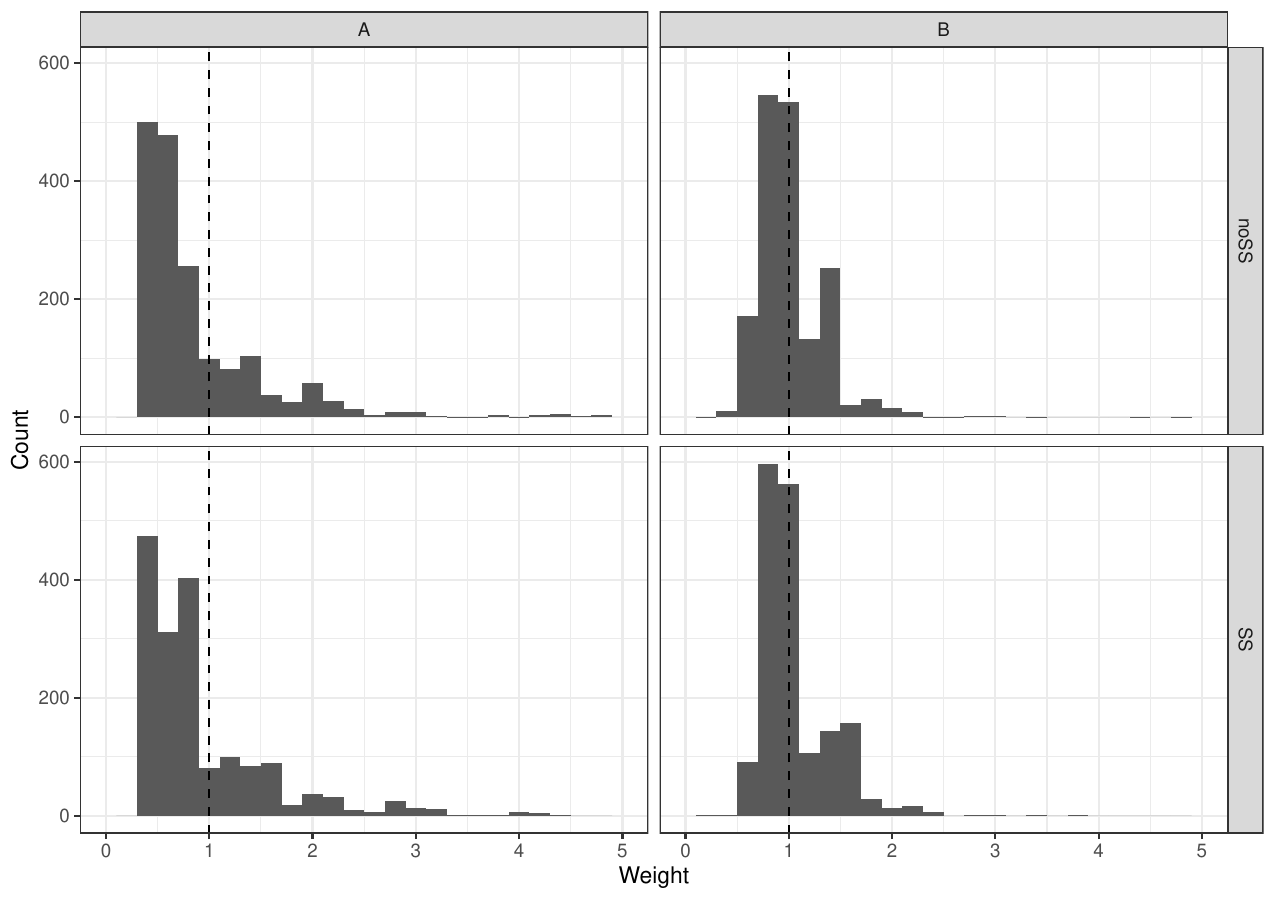}
    \caption{Least squares estimand weights approximated using the location-scale procedure described in \myappendix \ref{weight_approximations}. This procedure uses estimates of the conditional mean and variance or $A$ given $\bm{Z}$, which are obtained using the algorithms in Section \ref{proposed_algos} using the Super Learner for model fitting.}
    \label{fig:noniscrete-superlearner-weights}
\end{figure}

Finally, we examine how the least squares estimand weights in \eqref{ALSE-weight-new} looks in this example. Though this is not a necessary step for estimation of $\psi$ and $\Psi$, such approximations may give provide intuition as to how least squares estimands compare with average derivative estimands, i.e. where $w(a,\bm{z}) = 1$. We approximate weights under the location-scale exposure model of \cite{Klyne2023}, $A = \pi(\bm{Z}) + \beta^{1/2}(\bm{Z}) U$ where $U$ is a random variable with $U \indep \bm{Z}$. Details of the weight approximation procedure are in \myappendix \ref{weight_approximations}, with the resulting plots shown in Figure \ref{fig:noniscrete-superlearner-weights}. These plots highlight that least squares estimands are weighted average derivative estimands with, in this case, a modest reweighting of the observations.

\begin{table}[htb]
\caption{Least squares estimands applied to IWPC data. Results indicate the point estimates, its standard error, and 95\% Wald confidence interval, all in units of INR/(mg/week). P-values correspond to a Wald test of the null hypothesis that the estimand is zero.}
\centering
\begin{tabular}{|l|l|l|l|l|l|}
\hline
Estimand            & Algorithm                 & Estimate & SE & CI & p\\ \hline
$\Psi$   & noSS      & 1.98$\times10^{-3}$ & 6.58$\times10^{-4}$ &(0.692$\times10^{-3}$, 3.27$\times10^{-3}$) & 0.003    \\ \hline
$\Psi$   & SS      & 1.89$\times10^{-3}$ & 6.26$\times10^{-4}$ & (0.662$\times10^{-3}$, 3.12$\times10^{-3}$) & 0.003   \\ \hline
$\psi$   & noSS-A     & 0.260 & 0.200 & (-0.132, 0.651) & 0.19     \\ \hline
$\psi$   & SS-A     & 0.0910 & 0.126 & (-0.155, 0.337) & 0.47   \\ \hline
$\psi$   & noSS-B     & 1.57$\times10^{-3}$ & 8.40$\times10^{-4}$ & (-0.0748$\times10^{-3}$, 3.22$\times10^{-3}$) & 0.06 \\ \hline
$\psi$   & SS-B     & 1.34$\times10^{-3}$ & 9.46$\times10^{-4}$ & (-0.510$\times10^{-3}$, 3.20$\times10^{-3}$) & 0.15   \\ \hline
\end{tabular}
\label{IWPCresults}
\end{table}

\section{Related literature}
\label{related_lit}

\subsection{Causal Inference}
\label{sect:causal}

One of the main motivations to consider least squares estimands, and WADEs more generally, is to evaluate the main effect of a continuous exposure on an outcome. In recent work, \cite{Rothenhausler2019} propose a framework for so-called incremental treatment effects, and show that the ADE identifies their proposed estimand under limited causal assumptions. Here, we reframe their work to highlight the connection with WADEs.

Let $Y^a$ denote the outcome that would be observed if exposure had taken the value $A=a$. We define the conditional incremental effect at $a$ as
\begin{align*}
    \delta_{\epsilon}(a,\bm{z}) \equiv \frac{\E(Y^{a+\epsilon} - Y^a | \bm{Z}=\bm{z})}{\epsilon}
\end{align*}
which represents the scaled difference in conditional mean outcome in two counterfactual worlds, where all treatment units receive exposure level $a+\epsilon$ and $a$ respectively. We further define the weighted average incremental treatment effect $\Gamma_{w,\epsilon} \equiv \E\{w(A,\bm{Z})\delta_{\epsilon}(A,\bm{Z})\}$, and note that, for the weight $w(A,\bm{Z}) =1$, this estimand reduces to
\begin{align*}
\Gamma_{1,\epsilon} = \frac{\E(Y^{A+\epsilon}) - \E(Y^A)}{\epsilon}.
\end{align*}
Note that the introduction of the weight $w(a,\bm{z})$ represents a minor extension to the incremental treatment effects proposal.

One issue with incremental estimands of this type, is that it may be unclear which shift interventions $\epsilon$ should be considered, not least in the exploratory stage of an analysis where there may not be a particular intervention in mind. One option is to consider a curve such as, $\epsilon \mapsto \Gamma_{1,\epsilon}$, though there is no clear way to summarize the resulting curve once it has been obtained. Moreover, for large values of $\epsilon$, an exposure level of $A+\epsilon$ may be unrealistic for some treatment units, making the corresponding incremental treatment effect estimand $\Gamma_{1,\epsilon}$ scientifically uninteresting. Similar concerns also hold for so-called dose-response curves, i.e. the curve $a \mapsto \E(Y^a)$ which assigns the same exposure level for all treatment units \citep{Robins2001,Neugebauer2007,Kennedy2017}. 

In light of these concerns, \cite{Rothenhausler2019} propose a causal derivative estimand defined through the limit $\Gamma_{1,0} \equiv \lim_{\epsilon\to 0} \Gamma_{1,\epsilon}$, and they show that the ADE $\theta_1$ identifies $\Gamma_{1,0}$ assuming: (i) $Y^a\indep A|\bm{Z}$, (ii) $A=a \implies Y^a=Y$, (iii) continuity and differentiability of $a\mapsto f(a|\bm{z})$, (iv) $\lim_{\epsilon\to 0} \delta_{\epsilon}(a,\bm{z})$ is continuous and bounded. See proposition 1 of \cite{Rothenhausler2019} for details, which technically allows for a slightly weaker `local' version of (i). This recent connection between causal inference and the literature on nonparametric derivative estimands motivates a deeper understanding of WADEs.

\begin{remark}
Assumptions (i) and (ii) are also used to identify the WATE of a binary exposure on outcome, with (iii) replaced with the binary analogue, $P(A=a|\bm{Z}) > 0$ almost surely, for $a\in\{0,1\}$. The analogous identification result is $\E[w(\bm{Z})\{Y^1-Y^0\}] = \E[w(\bm{Z})\{\mu(1,\bm{Z}) - \mu(0,\bm{Z})\}]$.
\end{remark}

\subsection{Model projection}
\label{sect:projection}

Here we describe how least squares estimands in Example \ref{alse} are connected to least squares projection, and discuss other related observations. Consider a semi-parametric partially linear model, of the type studied by \cite{Robinson1988}, where the model, $\mathcal{M}_i$ is the set of functions of the form $\omega(\bm{z}) + \beta a$, indexed by the infinite dimensional parameter $(\beta,\omega)$, where $\omega: \mathbb{R}^p \mapsto \mathbb{R}$ is a function and $\beta \in \mathbb{R}$ is a constant.

Our goal is to find the model projection $\tilde{\mu}_{(i)}\in \mathcal{M}_i$ that is `nearest' to the unknown regression function $\mu(a,\bm{z})$ in the sense of minimising the norm, $||\mu - \tilde{\mu}||$. This notion of model projection is considered by \cite{Neugebauer2007} and \cite{Chambaz2012}, who propose projections on to similar linear working models, and also by \cite{Buja2019} who consider likelihood based projections. Projecting the regression function on to $\mathcal{M}_i$ gives,
\begin{align*}
\tilde{\mu}_{(i)}(a,\bm{z}) &\equiv \argmin_{g\in \mathcal{M}_i} \E \left[\{\mu(A,\bm{Z})-g(A,\bm{Z})\}^2 \right] \\
&=  \mu(\bm{z}) + \Psi \{a-\pi(\bm{z})\}.
\end{align*}
where $\mu(\bm{z})\equiv\E(Y|\bm{Z}=\bm{z})$. Hence we say the estimand $\Psi$ is a `least squares estimand' as it is the coefficient in a partially linear projection model which minimises the mean squared remainder. Crucially the model $\mathcal{M}_i$ is used to interpret the nonparametrically defined estimand $\Psi$, but we do not assume that the model is `true', in the sense that we do not require that $\mu \in \mathcal{M}_i$.

The projection view of least squares estimands is further extended by considering the (more flexible) conditionally linear working model $\mathcal{M}_{j} \supseteq \mathcal{M}_i$, which is the set of functions of the form $\omega(\bm{z}) + \nu(\bm{z}) a$, indexed by the infinite dimensional parameter $(\nu,\omega)$, where $\nu: \mathbb{R}^p \mapsto \mathbb{R}$ is a function. Projecting the regression function on to $\mathcal{M}_{j}$, as above gives,
\begin{align*}
\tilde{\mu}_{(j)}(a,\bm{z}) &\equiv \argmin_{g\in \mathcal{M}_{j}} \E \left[\{\mu(A,\bm{Z})-g(A,\bm{Z})\}^2 \right] \\
&=  \mu(\bm{z}) + \lambda(\bm{z}) \{a-\pi(\bm{z})\} \\
\lambda(\bm{z}) &\equiv \frac{\Cov(A,Y|\bm{Z}=\bm{z})}{\Var(A|\bm{Z}=\bm{z})}.
\end{align*}
Hence the effects described in Example \ref{alse} are `least squares estimands' since they represent weighted averages over the conditional least squares function $\lambda(\bm{z})$, i.e. they are of the form $\theta_w=\E\{w(\bm{Z})\lambda(\bm{Z})\}$, where $w(\bm{z}) = 1$ for $\psi$ and $w(\bm{z}) = \Var(A|\bm{Z} = \bm{z}) / \E\{\Var(A|\bm{Z})\}$ for $\Psi$. In both cases $w(\bm{Z})$ can be thought of as a normalised weight. 

\begin{remark}
    The function $\lambda(\bm{z})$ has particular relevance in the setting where $A\in\{0,1\}$ is a binary exposure, since, in that setting, $\mu \in\mathcal{M}_j$, and $\lambda(\bm{z})=\mu(1,\bm{z})-\mu(0,\bm{z})$ identifies the conditional ATE under standard causal assumptions.
\end{remark}

The estimand $\Psi$ also appears in \cite{Vansteelandt2020} who consider inference for the constant term indexing $h\{\mu(a,\bm{z})\}\in\mathcal{M}_i$ where $h(.)$ represents a canonical link function. Rather than consider model projection explicitly, they set out desirable properties of an estimand under model misspecification, defining a nonparametric estimand which reduces to $\Psi$, in the case of an identity link. Similarly, $\Psi$ appears elsewhere in the partially linear model setting without reference to projection \citep{Newey2018,Robins2008}.

The fact that least squares estimands are weighted ADEs is a novel contribution of this work, however, relates closely to three observations in the literature. The first, by \cite{Banerjee2007}, is that an estimator of the vector ADE may be constructed by partitioning the support of $\bm{X}$ into disjoint bins, and applying a linear regression to each bin. An ADE estimate is obtained by taking the average of these regression coefficients, weighted by the number of observations in each bin. The second observation, by \cite{Buja2019} is that the ordinary least squares (OLS) coefficient may be interpreted as a weighted sum of `slopes' between pairs of observations, without invoking differentiability. Thirdly, \cite{Hirshberg2018} show that when the response function is conditionally linear, i.e. $\mu \in\mathcal{M}_j$, then $\psi$ recovers the ADE. The key difference between \cite{Hirshberg2018} and the current work is that our interpretation does not rely on any functional form for $\mu(a,\bm{z})$ beyond differentiability, rather we interpret $\psi$ as an ADE with a certain kind of weighting.

\section{Discussion}

The current work makes several contributions to the literature on WADEs for a single covariate (exposure). In particular, we characterise the set of WADE RRs, allowing a broad set of estimands to be viewed as WADEs when the exposure is continuous and the conditional response function is differentiable. A causal interpretation of WADEs related to infinitesimal shift interventions means that a causal interpretation may be ascribed to all estimands of our class \citep{Rothenhausler2019}. Our RR set also characterises the RRs of WATEs, leading to a unified view of both estimands classes.

Next, we show that least squares estimands, which are estimands connected to partially linear model projections, are in fact WADEs for a particular choice of weight. We further motivate least squares estimands by considering the RR that minimises the nonparametric efficiency bound of the WADE, when the weight is known and the outcome is homoscedastic. Our efficiency analysis generalises the results of \cite{Crump2006} to the setting of continuous exposures.

Turning to estimation, we derive efficient one-step estimators, $\hat{\Psi}$ and $\hat{\psi}$, for least squares estimands, the latter of which generalises the AIPW to the setting of a continuous exposure. To estimate working models we recommend a quasi-oracle approach based on the R-learner \citep{Nie2017}, and a novel learner for the inverse variance, $1/\Var(A|\bm{Z})$, which we propose to mitigate extreme weighting in the estimator.

Our recommendation is that the least squares estimands $\psi$ and $\Psi$ be inferred for the exploratory analysis of the main effect of an exposure on an outcome. The resulting quantities may be interpreted as WADEs with data-adaptive weights. These weights happen to equal $1$ when the exposure is conditionally normally distributed.

\bibliography{refs}
\bibliographystyle{apalike}
\pagebreak
\appendix

\section{Weighted average derivative effect proofs}
\label{appendix:optimal_derivs}
\subsection{Proof of (\ref{canonical_contrast})}
\label{append:integration_by_parts}

Claim: Assume (C1), $\tilde{f}(a,\bm{z})\equiv w(a,\bm{z})f(a|\bm{z})$ is differentiable w.r.t. $a$, (C2), $\tilde{f}(s,\bm{z}) = \tilde{f}(t,\bm{z}) = 0$, where $s$ and $t$ denote the boundary of the support of $A$, and (C3) that $f(a|\bm{z})= 0$ implies $w(a,\bm{z})=0$. Then for any $h\in \mathcal{H}$, $\E\{w(A,\bm{Z})h^\prime(A,\bm{Z})\} = \E\{\alpha(A,\bm{Z})h(A,\bm{Z})\}$ with $\alpha$ given in $\eqref{canonical_contrast}$

Proof: Integration by parts gives
\begin{align*}
\E\{w(A,\bm{Z})h^\prime(A,\bm{Z})|\bm{Z} = \bm{z}\} &= \int_s^t  h^\prime(a,\bm{z}) \tilde{f}(a,\bm{z}) da \\
&=h(t,\bm{z})\tilde{f}(t,\bm{z})- h(s,\bm{z})\tilde{f}(s,\bm{z}) - \int_s^t  h(a,\bm{z}) \tilde{f}^\prime(a,\bm{z})  da \\
&= - \int_s^t  h(a,\bm{z}) \frac{\tilde{f}^\prime(a,\bm{z})}{f(a|\bm{z})} f(a|\bm{z}) da \\
&= \E\{ \alpha_w(A,\bm{Z})h(A,\bm{Z})|\bm{Z} = \bm{z}\}
\end{align*}
where we let $\alpha_w(a,\bm{z}) = -\tilde{f}^\prime(a,\bm{z}) / f(a|\bm{z})$, as in \eqref{canonical_contrast}. Using iterated expectations completes the proof.

Remark: To motivate Theorem \ref{alpha_theorem}, note that the fundamental theorem of calculus implies
\begin{align*}
\tilde{f}(a,\bm{z}) &= -\int_{s}^{a} \alpha_w(a^*,\bm{z})f(a^*| \bm{z}) da^* .
\end{align*}

\subsection{Proof of Theorem \ref{alpha_theorem}}
\label{main_theorem_proof}

First note that, the law of total expectation implies
\begin{align*}
    \E\{\alpha(A,\bm{Z})|\bm{Z}\} &= \E\{\alpha(A,\bm{Z})|A\leq a, \bm{Z}\}\{1 - F(a|\bm{Z})\} + \E\{\alpha(A,\bm{Z})|A > a, \bm{Z}\}F(a|\bm{Z}).
\end{align*}
Since $\E\{\alpha(A,\bm{Z})|\bm{Z}\} = 0$, for $\alpha \in \mathcal{R}$, we rewrite the weight in \eqref{f_tilde_theorem_alpha} as
\begin{align*}
    w(a,\bm{z})
    &= \frac{F(a|\bm{z})\{1-F(a|\bm{z})\}}{f(a|\bm{z})}\left[\E\{\alpha(A,\bm{Z})|A > a, \bm{Z}=\bm{z}\} - \E\{\alpha(A,\bm{Z})|A \leq a, \bm{Z}=\bm{z}\}\right] \\
    &= -\frac{\E\{\alpha(A,\bm{Z})|A\leq a,\bm{Z}=\bm{z}\}F(a|\bm{z}) - \E\{\alpha(A,\bm{Z})|\bm{Z}=\bm{z}\}F(a|\bm{z})}{f(a|\bm{z})} \\
    &= -\frac{\E\{\alpha(A,\bm{Z})|A\leq a,\bm{Z}=\bm{z}\}F(a|\bm{z}) }{f(a|\bm{z})} \\
    &= \frac{-1}{f(a|\bm{z})}\int_{s}^{a} \alpha(a^*,\bm{z})f(a^*| \bm{z}) da^* 
\end{align*}
where $s,t$ are the lower and upper boundary of the support of $A$. Therefore, for this weight, (C1) is satisfied by the fundamental theorem of calculus with \eqref{canonical_contrast} giving $\alpha_w = \alpha$. (C2) is satisfied due to the factor $F(a|\bm{z})\{1-F(a|\bm{z})\}$ which is zero for $a$ on the boundary of the support of $A$. (C3) is satisfied by the restrictions on $A$ in Theorem \ref{alpha_theorem}. We must therefore verify that $\E\{w(A,\bm{Z})\} \neq 0$ for any $\alpha \in \mathbb{R}$. We write
\begin{align}
  w(a,\bm{z})f(a|\bm{z}) &= - \int_s^t \alpha(a^*,\bm{z})\mathbb{I}(a^* \leq a) f(a^*|\bm{z}) da^*  \label{weight_integral}
\end{align}
where $\mathbb{I}(a^* \leq a)$ is 1 when $a^* \leq a$ and 0 otherwise. It follows that
\begin{align*}
  \E\left\{w(A,\bm{Z}) | \bm{Z} = \bm{z} \right\} &= \int_s^t w(a,\bm{z})f(a|\bm{z}) da \\
  &= -\int_s^t \int_s^t \alpha(a^*,\bm{z})\mathbb{I}(a^* \leq a) f(a^*|\bm{z}) da^* da \\
  &= \int_s^t \alpha(a^*,\bm{z}) \left[\int_s^t -\mathbb{I}(a^* \leq a) da \right] f(a^*|\bm{z}) da^* \\
  &= \int_s^t \alpha(a^*,\bm{z}) \left[ a^* - t \right] f(a^*|\bm{z}) da^* \\
  &= \E\{\alpha(A,\bm{Z})A|\bm{Z=z}\} - t \E\{\alpha(A,\bm{Z})|\bm{Z} = \bm{z}\} \\
  &= \E\{\alpha(A,\bm{Z})A|\bm{Z=z}\}
\end{align*}
hence $\E\left\{w(A,\bm{Z}) \right\} = \E\left\{ \alpha(A,\bm{Z})A \right\} = 1$.

\subsection{Proof of Lemma \ref{sufficiency}}
\label{sufficiency_proof}

Begin by splitting $\alpha(a,\bm{z})$ into a positive and negative part by defining two non-negative functions, $\alpha^{+}(a,\bm{z}) = \max \{\alpha(a,\bm{z}),0\}$ and $\alpha^{-}(a,\bm{z}) = \max \{-\alpha(a,\bm{z}),0\}$ such that, $\alpha(a,\bm{z}) = \alpha^{+}(a,\bm{z}) - \alpha^{-}(a,\bm{z})$. Since $\E\{\alpha(A,\bm{Z})|\bm{Z}\} = 0$, it follows that
\begin{align*}
\E\{\alpha^{+}(A,\bm{Z})|\bm{Z}\}&=\E\{\alpha^{-}(A,\bm{Z})|\bm{Z}\}.
\end{align*}
This equality is trivially satisfied by $\alpha^{+}(a|\bm{z})=\alpha^{-}(a|\bm{z}) = 0$, however this solution violates the requirement that $\E\{\alpha(A,\bm{Z})A\} =1 $, hence the positive and negative parts are both non-zero. Since, $\alpha(a, \bm{z})$ is monotonically increasing there must be some value, $c=c(\bm{z})$, on the support of $A$, such that the positive part is zero for $a<c$ and the negative part is zero for $a\geq c$, i.e.
\begin{align*}
\alpha(a,\bm{z}) &= - \alpha^{-}(a,\bm{z}) \mathbb{I}(a < c)  + \alpha^{+}(a, \bm{z}) \mathbb{I}(a \geq c).
\end{align*}
Next consider the quantity $\E\{\alpha(A,\bm{Z}) | A \leq a, \bm{Z} = \bm{z}\}F(a|\bm{z})$ when $a<c$,
\begin{align*}
\E\{\alpha(A,\bm{Z}) | A \leq a, \bm{Z} = \bm{z}\}F(a|\bm{z}) &= \int_s^a \alpha(a^*,\bm{z}) f(a^*|\bm{z}) da \\
&= - \int_s^a \alpha^{-}(a^*,\bm{z}) f(a^*|\bm{z}) da \leq 0.
\end{align*}
since $\alpha^{-}(a^*,\bm{z})$ is always positive. When $a\geq c$,
\begin{align*}
\E\{\alpha(A,\bm{Z}) | A \leq a, \bm{Z} = \bm{z}\}F(a|\bm{z})  &= \int_s^a \alpha(a^*,\bm{z}) f(a^*|\bm{z}) da   \\
&= \int_c^a \alpha^{+}(a^*,\bm{z}) f(a^*|\bm{z}) da  - \int_s^c \alpha^{-}(a^*,\bm{z})  f(a^*|\bm{z}) da. \\
&= \int_c^a \alpha^{+}(a^*,\bm{z}) f(a^*|\bm{z}) da  - \int_s^t \alpha^{+}(a^*,\bm{z})  f(a^*|\bm{z}) da. \\
&= - \int_a^t \alpha^{+}(a^*,\bm{z})  f(a^*|\bm{z}) da \leq 0
\end{align*}
Therefore, $\E\{\alpha(A,\bm{Z}) | A \leq a, \bm{Z} = \bm{z}\}F(a|\bm{z})  \leq 0$ for all $a$. Hence the weight in \eqref{f_tilde_theorem_alpha} is non-negative.

\subsection{Proof of Theorem \ref{optimality_1}}
\label{optimality_proof}

Background: An efficient estimator $\hat{\theta}_w $ of $\theta_w $ is regular asymptotically linear, such that
\begin{align*}
\hat{\theta}_w = \theta_w + n^{-1}\sum_{i=1}^n \phi_{\theta,w}(\bm{o}_i) + o_p(n^{-1/2})
\end{align*}
where $\phi_{\theta,w}(\bm{o})$ is the influence curve in \eqref{newey_IC}. Hence,
\begin{align*}
\sqrt{n}(\hat{\theta}_w - \theta_{w,S}) = n^{-1/2}\sum_{i=1}^n \alpha_w(a_i, \bm{z}_i)\{y_i-\mu(a_i,\bm{z}_i)\} + o_p(1)
\end{align*}
By the central limit theorem, $\sqrt{n}(\hat{\theta}_w - \theta_{w,S}) \overset{d}{\to}\n{0}{V}$, where the efficiency bound is
\begin{align*}
V &= \E\left\{\alpha_w^2(A,\bm{Z})\{Y-\mu(A,\bm{Z})\}^2\right\} \\
&= \E\left\{\alpha_w^2(A,\bm{Z})\sigma^2(A,\bm{Z})\right\} 
\end{align*}

Claim: $\E\left\{\alpha^2(A,\bm{Z})\sigma^2(A,\bm{Z})\right\}$ is minimised over $\alpha \in \mathcal{R}$ by the RR in Theorem \ref{optimality_1}.

Proof: Let $\gamma \in \mathcal{H}$ be a function with conditional mean $\bar{\gamma}(\bm{z}) \equiv \E\{\gamma(A,\bm{Z})|\bm{Z} = \bm{z}\}$. The function $\alpha(a,z) = \gamma(a,\bm{z}) - \bar{\gamma}(\bm{z})$ satisfies the condition $\E\{\alpha(A,\bm{Z})|\bm{Z}\} = 0$, and hence, $\alpha \in \mathcal{R}$ if and only if $\E\{\alpha(A,\bm{Z}) A \} = \E[\alpha(A,\bm{Z}) \{A - \pi(\bm{Z})\}] = 1$. We therefore consider the Lagrangian
\begin{align*}
  L(\gamma, \lambda) = \E\left( \{\gamma(A,\bm{Z}) - \bar{\gamma}(\bm{Z})\}^2 \sigma^2(A, \bm{Z}) -2\lambda [\{\gamma(A,\bm{Z}) - \bar{\gamma}(\bm{Z})\} \{A-\pi(\bm{Z})\} -1 ] \right)
\end{align*}
where $\lambda$ is a Lagrange multiplier.
Next consider replacing $\gamma(A,\bm{Z})$ with an alternative $\gamma(A,\bm{Z}) + \delta \eta(A,\bm{Z})$, where $\delta >0$ is constant. In doing so we obtain,
\begin{align*}
  \frac{L(\gamma + \delta \eta, \lambda) - L(\gamma, \lambda)}{\delta} = 2\E\left( \{\eta(A,\bm{Z}) - \bar{\eta}(\bm{Z})\}\left[\{\gamma(A,\bm{Z}) - \bar{\gamma}(\bm{Z})\}\sigma^2(A, \bm{Z}) -\lambda \{A-\pi(\bm{Z})\}\right] \right) + O(\delta)
\end{align*}
where $\bar{\eta}(\bm{z}) = \E\{\eta(A,\bm{Z})|\bm{Z}=\bm{z}\}$. We apply the identity
\begin{align*}
  \E\left(  \bar{\eta}(\bm{Z})\left[\{\gamma(A,\bm{Z}) - \bar{\gamma}(\bm{Z})\}\sigma^2(A, \bm{Z}) -\lambda \{A-\pi(\bm{Z})\}\right] \right) &= \E\left[\eta(A,\bm{Z}) \Cov\{\gamma(A,\bm{Z}), \sigma^2(A,\bm{Z})|\bm{Z}\}\right]
\end{align*}
To write
\begin{align*}
  \frac{L(\gamma + \delta \eta, \lambda) - L(\gamma, \lambda)}{\delta} = 2\E\left\{\eta(A,\bm{Z}) l(A,\bm{Z}, \gamma, \lambda) \right\} + O(\delta)
\end{align*}
where
\begin{align*}
  l(a,\bm{z}, \gamma, \lambda) &= \{\gamma(a,\bm{z}) - \bar{\gamma}(\bm{z})\}\sigma^2(a, \bm{z}) -\lambda \{a-\pi(\bm{z})\} - \Cov\{\gamma(A,\bm{Z}), \sigma^2(A,\bm{Z})|\bm{Z}=\bm{z}\}.
\end{align*}
By the fundamental lemma of the calculus of variations, we require that $l(a,\bm{z}, \gamma, \lambda) = 0$, and hence, writing $\alpha(a,\bm{z}) = \gamma(a,\bm{z}) - \bar{\gamma}(\bm{z})$ we require that
\begin{align*}
  \alpha(a,\bm{z})\sigma^2(a, \bm{z}) - \E\left[\alpha(A,\bm{Z})\sigma^2(A, \bm{Z}) | \bm{Z}=\bm{z} \right] = \lambda \{a - \pi(\bm{z})\}
\end{align*}
This equality is satisfied by
\begin{align*}
  \alpha(a, \bm{z}) &= \frac{\lambda \{a - \tilde{\pi}(\bm{z})\}}{\sigma^2(a, \bm{z})}.
\end{align*}
where the constraint $\E\{\alpha(A,\bm{Z}) | \bm{Z}\} = 0$ sets
\begin{align*}
  \tilde{\pi}(\bm{z}) \equiv \frac{\E\{A / \sigma^2(A,\bm{Z}) | \bm{Z} = \bm{z}\}}{\E\{1 / \sigma^2(A,\bm{Z}) | \bm{Z} = \bm{z}\}}.
\end{align*}
and the constraint $\E\{\alpha(A,\bm{Z}) A\} = 1$ sets
\begin{align*}
  \lambda  &= \E\left[\frac{\{A - \tilde{\pi}(\bm{Z})\}A}{\sigma^2(A, \bm{Z})} \right]^{-1}.
\end{align*}
which completes the proof.

\subsection{Theorem \ref{optimality_1} when the exposure is binary}
\label{crump_appendix}

Here we connect Theorem \ref{optimality_1} to Theorem 5.4 of \cite{Crump2006}, the latter of which considers the setting where $A \in \{0,1\}$ is a binary exposure. Letting $\sigma^2_a(\bm{z}) = \sigma^2(a,\bm{z})$ we obtain
\begin{align*}
  \tilde{\pi}(\bm{z}) &= \frac{\pi(\bm{z})\sigma^2_0(\bm{z})}{\pi(\bm{z})\sigma^2_0(\bm{z}) + \{1 - \pi(\bm{z})\}\sigma^2_1(\bm{z})}.
\end{align*}
Hence, by checking the $a=1,0$ cases separately, one can verify that
\begin{align*}
  \frac{a - \tilde{\pi}(\bm{z})}{\sigma^2(a,\bm{z})} &= \frac{a - \pi(\bm{z})}{\pi(\bm{z})\sigma^2_0(\bm{z}) + \{1 - \pi(\bm{z})\}\sigma^2_1(\bm{z})}.
\end{align*}
Following the definition in Theorem 5.4 of \cite{Crump2006}, we define the unnormalised weight
\begin{align*}
  w^*(\bm{z}) &\equiv \left(\frac{\sigma^2_1(\bm{z})}{\pi(\bm{z})} + \frac{\sigma^2_0(\bm{z})}{1 - \pi(\bm{z})} \right)^{-1} \\
  &= \frac{\pi(\bm{z}) \{1 - \pi(\bm{z})\}}{\pi(\bm{z})\sigma^2_0(\bm{z}) + \{1 - \pi(\bm{z})\}\sigma^2_1(\bm{z})}
\end{align*}
Using this weight we write
\begin{align*}
  \frac{a - \tilde{\pi}(\bm{z})}{\sigma^2(a,\bm{z})} = \frac{w^*(\bm{z}) \{a - \pi(\bm{z})\} }{\pi(\bm{z}) \{1 - \pi(\bm{z})\}}
\end{align*}
Thus we obtain the normalisation constant
\begin{align*}
  \E\left\{ \frac{\{A - \tilde{\pi}(\bm{Z})\} A }{\sigma^2(A,\bm{Z}) } \right\} &= \E\left\{ w^*(\bm{Z}) \frac{ \E[\{A - \pi(\bm{Z})\} A| \bm{Z}] }{\pi(\bm{Z}) \{1 - \pi(\bm{Z})\}} \right\} \\
  &= \E\{w^*(\bm{Z})\}.
\end{align*}
Hence the RR in Theorem \ref{optimality_1} reduces to 
\begin{align*}
  \alpha_w(a,\bm{z}) &= \left(\frac{w^*(\bm{z})}{\E\{w^*(\bm{Z})\}} \right) \frac{\{a - \pi(\bm{z})\} }{\pi(\bm{z}) \{1 - \pi(\bm{z})\}}
\end{align*}
which delivers the optimally weighted WATE estimand
\begin{align*}
  \langle \mu, \alpha_w \rangle &= \frac{\E[w^*(\bm{Z}) \{ \mu(1,\bm{Z}) - \mu(0,\bm{Z})\}] }{\E\{w^*(\bm{Z})\}}.
\end{align*}

\subsection{Proof of Lemma \ref{R_identLemma}}

Let $u(w)$ be a function such that $u(w)=0$ for $w= 0 $ and $u(w)=1/w$ otherwise. Consider the expectation
\begin{align*}
\E\left[ W\left\{Vu(W) - g(\bm{Z}) \right\}^2 \right] &= \E\left\{ V^2u^2(W)W \right\} + \E\left\{ g^2(\bm{Z})W - 2g(\bm{Z})VWu(W)\right\} \\
&= \E\left\{ V^2u^2(W)W \right\} + \E\left\{ g^2(\bm{Z})\E(W|\bm{Z}) - 2g(\bm{Z})\E\{VWu(W)|\bm{Z}\} \right\}
\end{align*}
For the purposes of minimization over $g$ the first term on the right hand side can be discarded since it does not depend on $g$. Hence
\begin{align*}
\argmin_{g \in \mathcal{F}} \E\left[ W\left\{Vu(W) - g(\bm{Z}) \right\}^2 \right] &= \argmin_{g \in \mathcal{F}} \E\left\{ g^2(\bm{Z})\E(W|\bm{Z}) - 2g(\bm{Z})\E\{VWu(W)|\bm{Z}\} \right\}
\end{align*}
By the calculus of variations, the minimiser $g^*(\bm{z})$ satisfies,
\begin{align*}
g^*(\bm{z})\E(W|\bm{Z}=\bm{z})  = \E\{VWu(W)|\bm{Z}=\bm{z}\}
\end{align*}
Since $W\neq 0 $ almost surely,
\begin{align*}
\E\{VWu(W)|\bm{Z}\} &= \E\{VWu(W)|W\neq0,\bm{Z}\}\\
&= \E(V|W\neq0,\bm{Z})\\
&= \E(V|\bm{Z}).
\end{align*}
Hence the result follows provided that $\E(W|\bm{Z}) \neq 0$ which is true since $W>0$ almost surely.

\section{Investigation into the Least squares estimand weight}

\subsection{Parametric distributions}
\label{append_ls_weight}

Here we examine the WADE weight $w_\psi(a, \bm{z})$ in Example \ref{alse}, which is associated with the least squares estimands, $\psi$ and proportional to the weight $w_\Psi(a,\bm{z}$ for $\Psi$ in the main text. We remark that this weight depends on the unknown distribution $P_0$ through the distribution of $A$ given $Z$, like the density weight in Example \ref{dens_ade} of the main text. The WADE weight for least squares estimands is therefore data-adaptive, though it is not necessary to compute this weight to estimate $\psi$ (or $\Psi$).

We consider the form of this exposure weight under various parametric distributions $f(a|\bm{z})$ in Pearson's distribution family. Table \ref{weight_table} summarises the functional form of the weight, with details provided in Examples \ref{norm_ex} to \ref{t_ex} below. For each distribution in Table \ref{weight_table}, we report the ``conditionally unnormalised'' weight, which recovers the true weight as $w_\psi(a, \bm{z}) = \tilde{w}(a, \bm{z}) / \E\{\tilde{w}(A, \bm{Z})|\bm{Z}=\bm{z}\}$.

\begin{table}[htbp]
\caption{Conditionally unnormalised weight $\tilde{w}(a, \bm{z})$ for least squares estimands when the distribution of $A$ given $Z$ follows a parametric distribution. Proofs below.}
\label{weight_table}
\centering
\begin{tabular}{|l|l|l|}
\hline
Exposure Distribution & Support & $\tilde{w}(a, z)$ \\ \hline
Normal & $A\in(-\infty, \infty)$ & $1$ \\
Gamma & $A>0$ & $a$ \\
Inverse-Gamma & $A>0$ & $a^2$ \\
Beta & $A\in(0,1)$ & $a(1-a)$ \\
Beta Prime & $A>0$ & $a(1+a)$ \\
t-distribution (d.o.f. $=\nu(\bm{z})$) & $A\in(-\infty, \infty)$ & $1+\frac{a^2}{\nu(\bm{z})}$ \\ \hline
\end{tabular}
\end{table}



When deriving the results in Table \ref{weight_table}, it is helpful to consider the function $\tilde{f}(a|\bm{z}) \equiv w_\psi(a, \bm{z})f(a|\bm{z})$. Since the weight is non-negative (by Lemma \ref{sufficiency}), and $\E\{w_\psi(A, \bm{Z})|\bm{Z}\} = 1$, it follows that $\tilde{f}(a|\bm{z})$ is a density function. Letting $s$ denote the lower boundary of the support of $A$ then we write this density function as
\begin{align*}
    \tilde{f}(a|\bm{z}) = \int_{s}^{a} \frac{\mu(\bm{z}) - a^*}{\beta(\bm{z})} f(a^*|\bm{z}) da^*
\end{align*}
We use this function as a tool to derive the exposure weight in the following examples. Each example requires verifying a derivative result, which follows from standard calculus and the Gamma function property $\nu\Gamma(\nu) = \Gamma(\nu + 1)$.

\begin{example}[Normal distribution]
\label{norm_ex}
Letting $f(a|\mu, \beta)$ denote the normal distribution, with mean $\mu = \mu(\bm{z})$ and variance $\beta = \beta(\bm{z})$, we claim that $\tilde{f}(a|\mu, \beta) = f(a|\mu, \beta)$, which implies a weight $w(a, \bm{z}) = \tilde{f}(a|\mu, \beta) / f(a|\mu, \beta) = 1$. To verify this claim, note that
\begin{align*}
\frac{d}{da} f(a|\mu, \beta) = \frac{\mu-a}{\beta} f(a|\mu, \beta).
\end{align*}
Hence, 
\begin{align*}
    \tilde{f}(a|\bm{z}) = \int_{-\infty}^{a} \left\{\frac{d}{da^*} f(a^*|\mu, \beta)\right\} da^*
\end{align*}
and the result follows by the fundamental theorem of calculus. The unitary weight of the normal distribution implies that the least squares estimand $\psi$ recovers the average derivative effect when the exposure $A$ is normally distributed given $Z$.
\end{example}

\begin{example}[Gamma distribution]
\label{gamma_ex}
The Gamma distribution, with shape parameter, $\nu = \nu(\bm{z}) > 0$, and rate parameter $\tau = \tau(\bm{z}) > 0$, has the density
\begin{align*}
f(a|\nu,\tau) = \frac{\tau^\nu}{\Gamma(\nu)} a^{\nu-1}\exp(-\tau a)
\end{align*}
for $a > 0$ and 0 otherwise. 

We claim that, for the gamma distribution, $\tilde{f}(a|\nu,\tau) = f(a|\nu+1,\tau)$. As in Example \ref{norm_ex}, it is sufficient to verify that
\begin{align*}
\frac{d}{da} f(a|\nu+1,\tau) = \frac{\mu-a}{\beta} f(a|\nu,\tau)
\end{align*}
where the mean and variance are $\mu=\nu/\tau$ and $\beta = \nu/\tau^2$. Therefore, the weight is
\begin{align*}
    w_\psi(a, \bm{z}) = \frac{f(a|\nu+1,\tau)}{f(a|\nu,\tau)} = \frac{a}{\mu}.
\end{align*}
This linear weight assigns unitary weight to the mean value $a=\mu$, with larger weight given to values above the mean.
\end{example}

\begin{example}[Inverse Gamma distribution]
\label{inv_gamma_ex}
The Inverse Gamma distribution, with shape parameter, $\nu = \nu(\bm{z}) > 0$, and scale parameter $\tau = \tau(\bm{z}) > 0$, has the density
\begin{align*}
f(a|\nu,\tau) = \frac{\tau^\nu}{\Gamma(\nu)} a^{-\nu-1}\exp(-\tau / a)
\end{align*}
for $a > 0$ and 0 otherwise. We claim that, for the inverse-gamma distribution, $\tilde{f}(a|\nu,\tau) = f(a|\nu-2,\tau)$. As before, it is sufficient to verify that
\begin{align*}
\frac{d}{da} f(a|\nu-2,\tau) = \frac{\mu-a}{\beta} f(a|\nu,\tau)
\end{align*}
where the mean and variance are $\mu = \tau/(\nu-1)$ and
\begin{align*}
    \beta = \frac{\tau^2}{(\nu-1)^{2}(\nu-2)}.
\end{align*}
Therefore, the weight is
\begin{align*}
    w_\psi(a, \bm{z}) = \frac{f(a|\nu-2,\tau)}{f(a|\nu,\tau)} = \frac{a^2}{\beta+\mu^2}.
\end{align*}
\end{example}

\begin{example}[Beta distribution]
\label{beta_ex}
The beta distribution, with shape parameters, $\nu = \nu(\bm{z}) > 0$, and $\tau = \tau(\bm{z}) > 0$, has the density
\begin{align*}
f(a|\nu,\tau) = \frac{\Gamma(\nu + \tau)}{\Gamma(\nu)\Gamma(\tau)} a^{\nu-1}(1-a)^{\tau-1}
\end{align*}
for $a \in [0,1]$ and 0 otherwise. We claim that, for the beta distribution, $\tilde{f}(a|\nu,\tau) = f(a|\nu+1,\tau+1)$. As before, it is sufficient to verify that
\begin{align*}
\frac{d}{da} f(a|\nu+1,\tau+1) = \frac{\mu-a}{\beta} f(a|\nu,\tau)
\end{align*}
where the mean and variance are $\mu = \nu / (\nu + \tau)$ and
\begin{align*}
    \beta = \frac{\nu\tau}{(\nu+\tau)^{2}(\nu+\tau+1)}.
\end{align*}
Therefore, the weight is
\begin{align*}
    w_\psi(a, \bm{z}) = \frac{f(a|\nu+1,\tau+1)}{f(a|\nu,\tau)} = \frac{a(1-a)}{\mu(1-\mu) - \beta}.
\end{align*}
This quadratic weight is at a maximum when $a=1/2$, with very little weight assigned to values close to $a=0,1$.
\end{example}

\begin{example}[Beta-prime distribution]
\label{beta_prime_ex}
The beta-prime distribution, with shape parameters, $\nu = \nu(\bm{z}) > 2$, and $\tau = \tau(\bm{z}) > 0$, has the density
\begin{align*}
f(a|\nu,\tau) = \frac{\Gamma(\nu + \tau)}{\Gamma(\nu)\Gamma(\tau)} a^{\nu-1}(1+a)^{-\nu-\tau}
\end{align*}
for $a \geq 0 $. We claim that, for the beta-prime distribution, $\tilde{f}(a|\nu,\tau) = f(a|\nu+1,\tau-2)$. As before, it is sufficient to verify that
\begin{align*}
\frac{d}{da} f(a|\nu+1,\tau-2) = \frac{\mu-a}{\beta} f(a|\nu,\tau)
\end{align*}
where the mean and variance are $\mu = \nu / (\tau - 1)$ and 
\begin{align*}
    \beta = \frac{\nu(\nu + \tau - 1)}{(\tau - 2)(\tau - 1)^2}.
\end{align*}
Therefore, the weight is
\begin{align*}
    w_\psi(a, \bm{z}) = \frac{f(a|\nu+1,\tau-2)}{f(a|\nu,\tau)} = \frac{a(1+a)}{\mu(1+\mu) + \beta}.
\end{align*}

\end{example}

\begin{example}[Student's t-distribution]
\label{t_ex}
The t-distribution, with $\nu = \nu(\bm{z}) > 2$ degrees of freedom, location parameter $\mu=\mu(\bm{z})$, and scale parameter $\tau = \tau(\bm{z})$ has the density
\begin{align*}
f(a|\nu, \mu, \tau^2) = \frac{\Gamma\left(\frac{\nu + 1}{2}\right)}{\sqrt{\nu\pi}\tau\Gamma\left(\frac{\nu}{2}\right)} \left(1 + \frac{(a-\mu)^2}{\nu \tau^2}\right)^{-\frac{\nu + 1}{2}}
\end{align*}
We claim that, for the t-distribution,
\begin{align*}
\tilde{f}(a|\nu, \mu, \tau^2) = f(a|\nu - 2, \mu, \beta)
\end{align*}
where $\beta = \tau^2\nu/(\nu-2)$ is the variance of the t-distribution. As before, it is sufficient to verify that
\begin{align*}
\frac{d}{da} f(a|\nu - 2, \mu, \beta) = \frac{\mu-a}{\beta} f(a|\nu, \mu, \tau^2).
\end{align*}
Therefore, the weight is
\begin{align*}
    w_\psi(a, \bm{z}) = \frac{f(a|\nu - 2, \mu, \beta)}{f(a|\nu, \mu, \tau^2)} = \frac{\left(1+\frac{(a-\mu)^2}{\nu\tau^2}\right)}{\left(1+\frac{1}{\nu-2}\right)}.
\end{align*}
As $\nu \to \infty$ then $w(a|\bm{z}) \to 1$. This is expected since the the t-distribution tends to a normal distribution, with mean $\mu$, and variance $\tau^2$ in this limit. Note that in Table \ref{weight_table}, this weight is reported for $\mu=0, \tau=1$.
\end{example}

\subsection{Numerical approximations}
\label{weight_approximations}

Here we propose a simple method to estimate least squares weights numerically. We emphasise that estimating weights is not necessary for inference or interpretation of least squares estimands, but may be useful for visually inspecting how, for a given dataset, least squares estimands might compare with an average derivative estimands, i.e. where $w(a,\bm{z}) = 1$. Following \cite{Klyne2023}, we assume a location-scale exposure model $A = \pi(\bm{Z}) + \beta^{1/2}(\bm{Z}) U$ where $U$ is a random variable with $U \indep \bm{Z}$. Under this model, we write the density
\begin{align*}
    f(a|\bm{z}) &= f_u \left\{ u(a,\bm{z})\right\} \\
    u(a,\bm{z}) &\equiv \frac{a - \pi(\bm{z})}{\beta^{1/2}(\bm{z})}
\end{align*}
where $f_u$ is the marginal density of $U$. Applying this density expression to the integral express of the least squares estimand weight in \eqref{weight_integral}, we obtain the weight
\begin{equation}
    w(a, \bm{z}) = \frac{-1}{f_u \{ u(a,\bm{z}) \}} \int_{-\infty}^\infty u^* \mathbb{I}\{u^* \leq u(a,\bm{z}) \} f_u(u^*) du^* \label{location-scale}
\end{equation}
where we have replaced the bounds of the support of $A$ with $\pm \infty$, and used the fact that, for least squares estimands, $\alpha(a,\bm{z}) =  u(a,\bm{z}) \beta^{-1/2}(\bm{z})$. To approximate \eqref{location-scale} numerically, we first estimate $u(a,\bm{z})$ by centering and scaling the observed exposure using estimates $\hat{\pi}$ and $\hat{\beta}$ that are obtained from the algorithms described in Section \ref{proposed_algos}. Next we estimate $f_u$ using a kernel density estimator, and approximate the integral in \eqref{location-scale} empirically using Monte-Carlo.

\section{Estimator Asymptotic Distribution}
\label{appendix:asymptotics}

In this Appendix we use a common empirical processes notation, where we define linear operators $P$ and $\mathbb{P}_n$ such that for some function $h(\bm{O})$, $P\{ h(\bm{O})\} \equiv \E \{h(\bm{O})\}$ and $\mathbb{P}_n \{h(\bm{O})\} \equiv n^{-1} \sum_{i=1}^n h(\bm{o}_i)$.

\subsection{Proof of Theorem \ref{asym_theorem_alse1}}
Define
\begin{align*}
\hat{\phi}_{\psi}(\bm{o}) &= \frac{\{y-\hat{\mu}(\bm{z})\}\{a-\hat{\pi}(\bm{z})\}-\hat{\lambda}(\bm{z})\{a-\hat{\pi}(\bm{z})\}^2}{\hat{\beta}(\bm{z})} + \hat{\lambda}(\bm{z}) -\hat{\psi}_0
\end{align*}
where $\hat{\psi}_0 = \mathbb{P}_n\{\hat{\lambda}(\bm{Z})\}$ denotes an initial estimate of $\psi$. Without making any restrictions we write
\begin{align}
    \hat{\psi} - \psi &= (\mathbb{P}_n -P)\{\phi_{\psi}(\bm{O})\} + R_n + H_n \\
    \hat{\psi} &\equiv \hat{\psi}_0 + \mathbb{P}_n\{\hat{\phi}_{\psi}(\bm{O})\} \\
    R_n &\equiv \hat{\psi}_0 - \psi + P\{\hat{\phi}_{\psi}(\bm{O})\} \\
    H_n &\equiv (\mathbb{P}_n -P)\{\hat{\phi}_{\psi}(\bm{O}) - \phi_{\psi}(\bm{O})\}.
\end{align}
We will show that the remainder therm $R_n=o_P(n^{-1/2})$ and the empirical process term $H_n=o_P(n^{-1/2})$, and hence the result follows since $P\{\phi_{\psi}(\bm{O})\} = 0$. 

\textbf{The remainder term}

To simplify notation, we will omit function arguments, e.g. $\mu = \mu(\bm{Z})$ with similar for $\hat{\mu},\lambda,\hat{\lambda},\pi,\hat{\pi},\hat{\beta},\beta$. Since $\psi = \E[\lambda]$ we can write the remainder term as
\begin{align*}
R_n &= \E\left[ \frac{(Y-\hat{\mu})(A-\hat{\pi})-\hat{\lambda}(A-\hat{\pi})^2}{\hat{\beta}} + \hat{\lambda}-\lambda \right]
\end{align*}
Note that
\begin{align*}
\E[(Y-\hat{\mu})(A-\hat{\pi})|\bm{Z}] &= \lambda\beta + (\mu-\hat{\mu})(\pi-\hat{\pi}) \\
\E[(A-\hat{\pi})^2|\bm{Z}] &= \beta + (\pi-\hat{\pi})^2
\end{align*}
and hence
\begin{align*}
R_n &= \E\left[ \frac{(\pi-\hat{\pi}) \{\mu-\hat{\mu}-\hat{\lambda}(\pi-\hat{\pi}) \} + (\lambda-\hat{\lambda})(\beta-\hat{\beta}) }{\hat{\beta}} \right] 
\end{align*}
Using the inequality, $(a+b)^2\leq 2(a^2+b^2)$,
\begin{align}
\frac{R_n^2}{2} &\leq \underbrace{\E\left[ \frac{(\pi-\hat{\pi}) \{\mu-\hat{\mu}-\hat{\lambda}(\pi-\hat{\pi}) \}}{\hat{\beta}} \right]^2}_{\text{first remainder}}  + \underbrace{\E\left[\frac{(\lambda-\hat{\lambda})(\beta-\hat{\beta}) }{\hat{\beta}} \right]^2 }_{\text{second remainder}} \label{appendix_remainder}
\end{align}
We will show that the two remainder terms on the right hand side are $o_p(n^{-1})$. For the first remainder, the Cauchy-Schwarz inequality gives
\begin{align*}
\E\left[ \frac{(\pi-\hat{\pi}) \{\mu-\hat{\mu}-\hat{\lambda}(\pi-\hat{\pi}) \}}{\hat{\beta}} \right]^2 &\leq \E\left[\frac{(\pi-\hat{\pi})^2}{\hat{\beta}}\right] \E\left[\frac{\{\mu-\hat{\mu}-\hat{\lambda}(\pi-\hat{\pi}) \}^2}{\hat{\beta}}\right] \\
&\leq \left(\frac{1}{\epsilon^2}\right) \E\left[(\pi-\hat{\pi})^2\right] \E\left[\{\mu-\hat{\mu}-\hat{\lambda}(\pi-\hat{\pi}) \}^2\right] \\
&\leq \left(\frac{2}{\epsilon^2}\right) \E\left[(\pi-\hat{\pi})^2\right] \left\{\E\left[(\mu-\hat{\mu})^2\right] + \E\left[\hat{\lambda}^2(\pi-\hat{\pi})^2\right] \right\} \\
&\leq \left(\frac{2}{\epsilon^2}\right) \E\left[(\pi-\hat{\pi})^2\right] \left\{\E\left[(\mu-\hat{\mu})^2\right] + K^2\E\left[(\pi-\hat{\pi})^2\right] \right\}
\end{align*}
To obtain the second inequality above, we choose $\epsilon > 0$ such that $\hat{\beta} \geq \epsilon$ almost surely, and to obtain the third inequality we once again apply the inequality $(a+b)^2\leq 2(a^2+b^2)$. It therefore follows from (A1) that the first remainder in \eqref{appendix_remainder} is $o_p(n^{-1})$.

For the second remainder in \eqref{appendix_remainder}, the Cauchy-Schwarz inequality gives
\begin{align*}
\E\left[\frac{(\lambda-\hat{\lambda})(\beta-\hat{\beta})}{\hat{\beta}}  \right]^2 &\leq \E\left[\frac{(\lambda-\hat{\lambda})^2}{\hat{\beta}}\right] \E\left[\frac{(\beta-\hat{\beta})^2}{\hat{\beta}}\right] \\
&\leq \left(\frac{1}{\epsilon^2}\right) \E\left[(\lambda-\hat{\lambda})^2\right] \E\left[(\beta-\hat{\beta})^2\right]
\end{align*}
which is $o_P(n^{-1})$ under (A2). Hence $R_n=o_P(n^{-1/2})$.

\textbf{The empirical process term}

First write the empirical process term as the sum of four terms
\begin{align*}
    H_n &= (\mathbb{P}_n - P)\{\psi - \hat{\psi}_0\} \\
    &+ (\mathbb{P}_n - P)\{\hat{\lambda} - \lambda\} \\
    &+ (\mathbb{P}_n - P)\left\{\frac{\hat{u}(\bm{O})}{\hat{\beta}} - \frac{u(\bm{O})}{\beta}\right\} \\
    &+ (\mathbb{P}_n - P)\left\{\frac{\hat{\lambda}\hat{v}(\bm{O})}{\hat{\beta}} - \frac{\lambda v(\bm{O})}{\beta}\right\}
\end{align*}
where
\begin{align*}
    \hat{u}(\bm{O}) &\equiv (Y-\hat{\mu})(A-\hat{\pi}) \\
    u(\bm{O}) &\equiv (Y-\mu)(A-\pi) \\
    \hat{v}(\bm{O}) &\equiv (A-\hat{\pi})^2 \\
    v(\bm{O}) &\equiv (A-\pi)^2
\end{align*}
Note that the first term is zero since $(\mathbb{P}_n - P)\{\psi - \hat{\psi}_0\} = (\psi - \hat{\psi}_0)(\mathbb{P}_n - P)\{1\} = 0$. When the Donsker condition holds, then, by Lemma 19.24 of \cite{Vaart2013}, the second term is $o_P(n^{-1/2})$ provided (i) that $\E\{(\hat{\lambda} - \lambda)^2\} = o_p(1)$, the third term is $o_P(n^{-1/2})$ provided (ii) that $\E\left[\left\{\frac{\hat{u}(\bm{O})}{\hat{\beta}} - \frac{u(\bm{O})}{\beta}\right\}^2\right] = o_p(1)$, and the fourth term is $o_P(n^{-1/2})$ provided (iii) that $\E\left[\left\{\frac{\hat{\lambda}\hat{v}(\bm{O})}{\hat{\beta}} - \frac{\lambda v(\bm{O})}{\beta}\right\}^2\right] = o_p(1)$. Similarly, under sample splitting then by Chebyshev's inequality, (i), (ii), and (iii) are also sufficient conditions for $H_n$ to be $o_P(n^{-1/2})$. Note that condition (i) holds since $\hat{\lambda}(\bm{z})$ is a consistent estimator of $\lambda(\bm{z})$, therefore, we must only show that (ii) and (iii) hold.

For (ii) we write
\begin{align*}
    \frac{\hat{u}(\bm{O})}{\hat{\beta}} - \frac{u(\bm{O})}{\beta} &= \frac{\hat{u}(\bm{O}) - u(\bm{O})}{\hat{\beta}} + \left\{\frac{1}{\hat{\beta}} - \frac{1}{\beta}\right\}u(\bm{O})  \\
    &= \frac{(Y-\mu) \epsilon_\pi + (A-\pi) \epsilon_\mu + \epsilon_\mu \epsilon_\pi}{\hat{\beta}} - \frac{\epsilon_\beta (Y-\mu)(A-\pi)}{\beta\hat{\beta}}
\end{align*}
where $\epsilon_\mu = \hat{\mu} - \mu$, $\epsilon_\pi = \hat{\pi} - \pi$, and $\epsilon_{\beta} = \hat{\beta} - \beta$. Hence, conditioning on the sample that delivered the nuisance parameter estimators (which we do not make explicit in our notation)
\begin{align*}
    \E\left[\left\{\frac{\hat{u}(\bm{O})}{\hat{\beta}} - \frac{u(\bm{O})}{\beta} \right\}^2 \Big| \bm{Z}\right] &= \frac{k_{2,0}(\bm{Z})\epsilon_\pi^2 + k_{0,2}(\bm{Z})\epsilon_\mu^2 + 2k_{1,1}(\bm{Z})\epsilon_\mu\epsilon_\pi + \epsilon^2_\mu \epsilon_\pi^2}{\hat{\beta}^2} \\
    &+ \frac{\epsilon_{\beta}^2k_{2,2}(\bm{Z})}{\beta^2\hat{\beta}^2} \\
    &+ \frac{2\epsilon_\beta}{\beta\hat{\beta}^2}\left\{k_{2,1}(\bm{Z})\epsilon_\pi + k_{1,2}(\bm{Z}) \epsilon_\mu + k_{1,1}(\bm{Z})\epsilon_\pi \epsilon_\mu \right\}
\end{align*}
where $k_{i,j}(\bm{Z}) = \E\{(Y-\mu)^i(A-\pi)^j|\bm{Z}\}$. By Cauchy-Schwarz, $k_{i,j}^2(\bm{Z}) \leq E\{(Y-\mu)^{2i}|\bm{Z}\}E\{(A-\pi)^{2j}|\bm{Z}\}$ hence each of the $k_{i,j}^2(\bm{Z}) \leq K^2$ in the expression above
\begin{align*}
   \Bigg| \E\left[\left\{\frac{\hat{u}(\bm{O})}{\hat{\beta}} - \frac{u(\bm{O})}{\beta} \right\}^2 \Big| \bm{Z}\right] \Bigg| \leq \frac{K}{\epsilon^2} \{\epsilon_{\pi}^2 + \epsilon_{\mu}^2 + 2 |\epsilon_\mu \epsilon_\pi|\} + \frac{\epsilon_{\pi}^2 \epsilon_{\mu}^2}{\epsilon^2} + \frac{K}{\epsilon^4} \epsilon_{\beta}^2 + \frac{2K}{\epsilon^3}|\{\epsilon_{\pi} + \epsilon_{\mu} + \epsilon_{\pi}\epsilon_{\mu}\}|
\end{align*}
Since each of $\pi,\mu,\beta$ are consistent, the right hand side above tends to zero, as does (ii) by the dominated convergence theorem. The proof of (iii) proceeds in a similar way,
\begin{align*}
    \frac{\hat{\lambda}\hat{v}(\bm{O})}{\hat{\beta}} - \frac{\lambda v(\bm{O})}{\beta} &= \frac{\hat{\lambda} \epsilon_{\pi} \{2(A-\pi) + \epsilon_{\pi} \} }{\hat{\beta}} + \frac{\{\epsilon_{\lambda} \beta - \lambda \epsilon_\beta\}(A-\pi)^2}{\beta \hat{\beta}} \\
    \E\left[\left\{\frac{\hat{\lambda}\hat{v}(\bm{O})}{\hat{\beta}} - \frac{\lambda v(\bm{O})}{\beta} \right\}^2 \Big| \bm{Z}\right] 
    &= \frac{\hat{\lambda}^2}{\hat{\beta}^2}\epsilon_\pi^2\{2\beta + \epsilon_\pi^2\} \\ 
    &+ \frac{\{\epsilon_{\lambda} \beta - \lambda \epsilon_\beta\}^2 \E\{(A-\pi)^4|\bm{Z}\}}{\beta^2\hat{\beta}^2} \\ 
    &+ \frac{2\hat{\lambda}\epsilon_\pi[2\E\{(A-\pi)^3|\bm{Z}\} + \epsilon_\pi \beta]\{\epsilon_{\lambda} \beta - \lambda \epsilon_\beta\}}{\beta\hat{\beta}^2}
\end{align*}
Note that Cauchy-Schwarz implies $\E\{(A-\pi)^3|\bm{Z}\}^2 \leq \E\{(A-\pi)^4|\bm{Z}\}\beta < K^2$, and $\lambda^2 \leq \beta \Var(Y|\bm{Z}) < K^2$.

\subsection{Proof of Theorem \ref{asym_theorem_alse2}}

Let
\begin{align*}
\gamma &\equiv P\left[ \{Y-\mu(\bm{Z})\}\{A-\pi(\bm{Z})\}\right] \\
\hat{\gamma} &\equiv \mathbb{P}_n\left[ \{Y-\hat{\mu}(\bm{Z})\}\{A-\hat{\pi}(\bm{Z})\}\right] \\
\eta &\equiv P\left[ \{A-\pi(\bm{Z})\}^2\right] \\
\hat{\eta} &\equiv \mathbb{P}_n\left[\{A-\hat{\pi}(\bm{Z})\}^2\right] \\
\phi_{\gamma}(\bm{O}) &\equiv \{Y-\mu(\bm{Z})\}\{A-\pi(\bm{Z})\} - \gamma \\
\hat{\phi}_{\gamma}(\bm{O}) &\equiv \{Y-\hat{\mu}(\bm{Z})\}\{A-\hat{\pi}(\bm{Z})\} - \hat{\gamma} \\
\phi_{\eta}(\bm{O}) &\equiv \{A-\pi(\bm{Z})\}^2 - \eta \\
\hat{\phi}_{\eta}(\bm{O}) &\equiv \{A-\hat{\pi}(\bm{Z})\}^2 - \hat{\eta}
\end{align*}
Without making any restrictions we write
\begin{align*}
    \hat{\gamma} - \gamma &= (\mathbb{P}_n -P)\{\phi_{\gamma}(\bm{O})\} + R_n + H_n \\
    R_n &\equiv \hat{\gamma} - \gamma + P\{\hat{\phi}_{\gamma}(\bm{O})\} \\
    H_n &\equiv (\mathbb{P}_n -P)\{\hat{\phi}_{\gamma}(\bm{O}) - \phi_{\gamma}(\bm{O})\}.
\end{align*}
We will show that the remainder therm $R_n=o_P(n^{-1/2})$ and the empirical process term $H_n=o_P(n^{-1/2})$. Since $P\{\phi_{\gamma}(\bm{O})\} = 0$, we therefore obtain the RAL results
\begin{align*}
    \hat{\gamma} - \gamma &= \mathbb{P}_n\{\phi_{\gamma}(\bm{O})\} + o_p(n^{-1/2}) \\
    \hat{\eta} - \eta &= \mathbb{P}_n\{\phi_{\eta}(\bm{O})\} + o_p(n^{-1/2})
\end{align*}
where the result for $\hat{\eta}$ follows as a special case ($Y=A$) of the result for $\hat{\gamma}$. Our goal is to consider the estimator $\hat{\Psi} = \hat{\gamma}/\hat{\eta}$, which estimates $\Psi = \gamma/\eta$. It follows by algebraic manipulations that,
\begin{align*}
\sqrt{n}(\hat{\Psi} - \Psi) &= \frac{\eta}{\hat{\eta}} \left[ \sqrt{n}\mathbb{P}_n \{\phi_{\Psi}(\bm{O})\} + o_p(1) \right]
\end{align*}
where $ \phi_{\Psi}(\bm{o})  = \{\phi_{\gamma}(\bm{o}) - \Psi \phi_{\eta}(\bm{o})\}/\eta$ is the IC of $\Psi$. Next we use Slutsky's Theorem and the fact that $\hat{\eta}/\eta$ converges to 1 in probability, to write,
\begin{align*}
\lim_{n\to \infty} \sqrt{n}(\hat{\Psi} - \Psi) &= \lim_{n\to \infty} \sqrt{n}\mathbb{P}_n \{\phi_{\Psi}(\bm{O})\}
\end{align*}
which is the desired result.

\textbf{The remainder term}
To simplify notation, we will omit function arguments, e.g. $\mu = \mu(\bm{Z})$ with similar for $\hat{\mu},\pi,\hat{\pi}$. Evaluating the remainder gives
\begin{align*}
R_n &= \E\left[ (Y-\hat{\mu})(A-\hat{\pi})-  (Y-\mu)(A-\pi) \right] \\
&= \E\left[ (\mu-\hat{\mu})(\pi-\hat{\pi})\right].
\end{align*}
By the Cauchy-Schwarz inequality $R_n^2 \leq \E\left[ (\mu-\hat{\mu})^2\right] \E\left[(\pi-\hat{\pi})^2\right]$, which is $o_P(n^{-1})$ under (A1).

\textbf{The empirical process term}

First write the empirical process term as the sum
\begin{align*}
    H_n &= (\mathbb{P}_n - P)\{\gamma - \hat{\gamma}\}\\
    &+ (\mathbb{P}_n - P)\left[\{Y-\mu(\bm{Z})\}\{\hat{\pi}(\bm{Z}) - \pi(\bm{Z})\}\right] \\
    &+ (\mathbb{P}_n - P)\left[\{\hat{\mu}(\bm{Z}) - \mu(\bm{Z})\}\{A-\pi(\bm{Z})\}\right] \\
    &+ (\mathbb{P}_n - P)\left[\{\hat{\mu}(\bm{Z}) - \mu(\bm{Z})\}\{\hat{\pi}(\bm{Z}) - \pi(\bm{Z})\}\right]
\end{align*}
Note that the first term is zero since $(\mathbb{P}_n - P)\{\gamma - \hat{\gamma}\} = (\gamma - \hat{\gamma})(\mathbb{P}_n - P)\{1\} = 0$. When the Donsker condition holds, then, by Lemma 19.24 of \cite{Vaart2013}, the remaining terms are $o_P(n^{-1/2})$ provided (i) that 
\begin{align*}
&\E\left[\Var(Y|\bm{Z})\{\hat{\pi}(\bm{Z}) - \pi(\bm{Z})\}^2\right] \\
&\E\left[\{\hat{\mu}(\bm{Z}) - \mu(\bm{Z})\}^2\Var(A|\bm{Z})\right] \\
&\E\left[\{\hat{\mu}(\bm{Z}) - \mu(\bm{Z})\}^2\{\hat{\pi}(\bm{Z}) - \pi(\bm{Z})\}^2\right]
\end{align*}
Similarly, under sample splitting then by Chebyshev's inequality, (i) is also sufficient conditions for $H_n$ to be $o_P(n^{-1/2})$. We claim that (i) holds since $\hat{\mu}$ and $ \hat{\pi}$ are consistent and $\Var(A|\bm{Z}) < K$ and $\Var(Y|\bm{Z}) < K$.

\section{Additional illustrated results}
\label{appendix:iwpc_extra}

\begin{table}[htb]
\caption{Least squares estimands applied to IWPC data, using the discrete Super Learner algorithm for model fitting. Results indicate the point estimates, its standard error, and 95\% Wald confidence interval, all in units of INR/(mg/week). P-values correspond to a Wald test of the null hypothesis that the estimand is zero.}
\centering
\begin{tabular}{|l|l|l|l|l|l|}
\hline
Estimand            & Algorithm                 & Estimate & SE & CI & p\\ \hline
$\Psi$     &noSS    & 1.87$\times10^{-3}$ & 6.32$\times10^{-4}$ &(0.633$\times10^{-3}$, 3.11$\times10^{-3}$) & 0.003       \\ \hline
$\Psi$     &SS    & 1.85$\times10^{-3}$ & 6.26$\times10^{-4}$ & (0.623$\times10^{-3}$, 3.08$\times10^{-3}$) & 0.003   \\ \hline
$\psi$     &noSS-A   & -0.105 & 0.956 & (-1.98, 1.77) & 0.91      \\ \hline
$\psi$     &SS-A   & -0.269 & 0.236 & (-0.732, 0.193) & 0.25   \\ \hline
$\psi$     &noSS-B   & 1.46$\times10^{-3}$ & 8.13$\times10^{-4}$ & (-0.133$\times10^{-3}$, 3.05$\times10^{-3}$) & 0.07       \\ \hline
$\psi$     &SS-B   & 1.37$\times10^{-3}$ & 8.43$\times10^{-4}$ & (-0.282$\times10^{-3}$, 3.02$\times10^{-3}$) & 0.10   \\ \hline
\end{tabular}
\label{IWPCresults2}
\end{table}

\begin{figure}[htb]
    \centering
    \includegraphics[width=\linewidth]{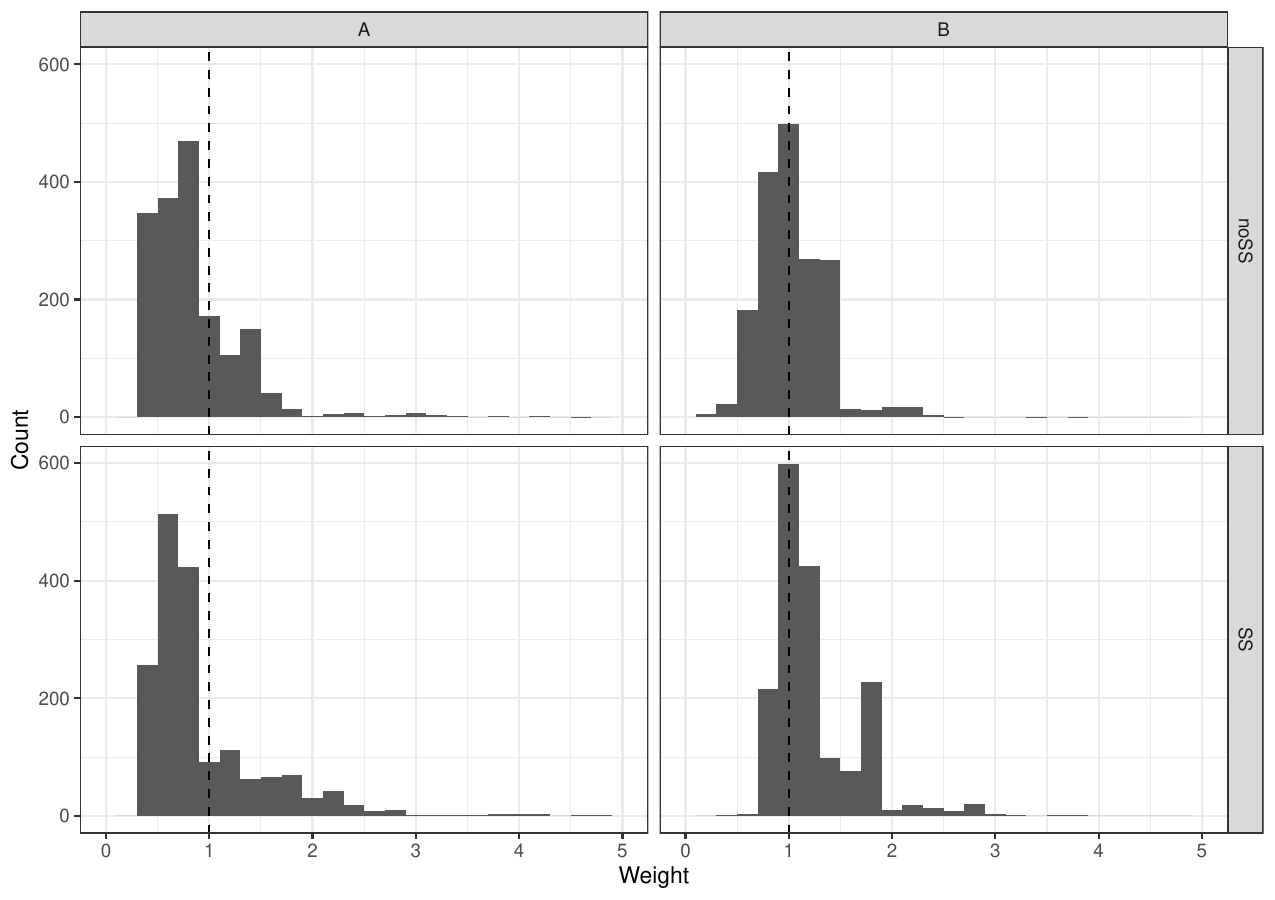}
    \caption{Least squares estimand weights approximated using the location-scale procedure described in Supplement \ref{weight_approximations}. This procedure uses estimates of the conditional mean and variance or $A$ given $\bm{Z}$, which are obtained using the algorithms in Section \ref{proposed_algos} and a discrete super learner for model fitting.}
    \label{fig:discrete-superlearner-weights}
\end{figure}
\end{document}